\documentclass[12pt]{article}
\usepackage{amsfonts}

\begin{document}
\renewcommand{\thefootnote}{\arabic{footnote}}
\newcommand{\fn}{\footnote}
\newcommand{\be}{\begin{equation}}
\newcommand{\bea}{\begin{eqnarray}}
\newcommand{\ee}{\end{equation}}
\newcommand{\eea}{\end{eqnarray}}
\newcommand{\sect}[1]{\setcounter{equation}{0}\bigskip\medskip\section{#1}\smallskip}
\newcommand{\subsect}[1]{\medskip\subsection{#1}\smallskip}
\newcommand{\subsubsect}[1]{\medskip\subsubsection{#1}\smallskip}
\renewcommand{\theequation}{\thesection.\arabic{equation}}
\newtheorem{proposition}{Proposition}[section]  
\newcommand{\bprop}{\medskip\begin{proposition}
\it}
\newcommand{\eprop}{\end{proposition} \hfill }
\newcommand{\proof}{\medskip \noindent \\{\it Proof.} \rm }
\newtheorem{naming}{Definition}[section]   
\newcommand{\bdefi}{\medskip\begin{naming}
\it}
\newcommand{\edefi}{\end{naming} \hfill }
\newtheorem{example}{Example}[section]   
\def\bexam{\medskip\begin{example}
\rm}
\def\eexam{\end{example} \hfill }
\def\norm#1{{\vert\vert#1\vert\vert}}
\def\abs#1{{\vert#1\vert}}
\def\ha{\widehat{\cal A}}
\def\hc{\widehat{\cal C}}
\def\prim{Prim{\cal A} }
\def\comp{{\cal K}({\cal H})}
\def\inf1{{\cal L}^{(1,\infty)}}
\def\omca{\Omega {\cal A}}
\def\oca#1{\Omega^{#1}{\cal A}}
\def\omcad{\Omega_D {\cal A}}
\def\ocad#1{\Omega_D^{#1}{\cal A}}
\def\ota{\otimes_{\cal A}}
\def\otc{\otimes_{\IC}}
\def\join{{\protect{\footnotesize $\backslash\!/~$}}}
\def\meet{{\protect{\footnotesize $/\!\backslash~$}}}
\def\bar#1{\overline{#1}}
\def\wh{\widehat}
\def\wt{\widetilde}
\def\bra#1{\left\langle #1\right|}
\def\ket#1{\left| #1\right\rangle}
\def\hs#1#2{\left\langle #1,#2\right\rangle}
\def\vev#1{\left\langle #1\right\rangle}
%
%
\def\sdp{\hbox{ \raisebox{.25ex}{\tiny $|$}\hspace{.2ex}{$\!\!\times $}} }
\def\pa{\partial}
\def\del{\nabla}
\def\ca{{\cal A}}
\def\cb{{\cal B}}
\def\cc{{\cal C}}
\def\cd{{\cal D}}
\def\ce{{\cal E}}
\def\cf{{\cal F}}
\def\cg{{\cal G}}
\def\ch{{\cal H}}
\def\ci{{\cal I}}
\def\cj{{\cal J}}
\def\ck{{\cal K}}
\def\cl{{\cal L}}
\def\cm{{\cal M}}
\def\cn{{\cal N}}
\def\co{{\cal O}}
\def\cp{{\cal P}}
\def\cs{{\cal S}}
\def\ct{{\cal T}}
\def\cu{{\cal U}}
\def\cv{{\cal V}}
\def\cw{{\cal W}}
\def\cx{{\cal X}}
\def\cuai{{\cal Y}}
\def\cz{{\cal Z}}
\def\bca{\overline{{\cal A}}}
\def\IC{{\mathbb C}}
\def\ID{{\mathbb D}}
\def\IF{{\mathbb F}}
\def\IH{{\mathbb H}}
\def\II{{\mathbb I}}
\def\I1{{\mathbb I}}
\def\IK{{\mathbb K}}
\def\IM{{\mathbb M}}
\def\IN{{\mathbb N}}
\def\IP{{\mathbb P}}
\def\IQ{{\mathbb Q}}
\def\IZ{{\mathbb Z}}
\def\IR{{\mathbb R}}
\def\otc{\otimes_{\IC}}
\def\otca{\otimes_{{\cal A}}}
\def\bc{{\bf C}}
\def\br{{\bf R}}
\def\bz{{\bf Z}}
\def\bn{{\bf N}}
\def\bm{{\bf M}}
\def\bt{{\bf T}}
\def\Up{\Uparrow}
\def\up{\uparrow}
\def\Dn{\Downarrow}
\def\dn{\downarrow}
\def\Ra{\Rightarrow}
\def\ra{\rightarrow}
\def\La{\Leftarrow}
\def\la{\leftarrow}
\def\lra{\longrightarrow}
\def\iff{\Leftrightarrow}
\thispagestyle{empty}
\setcounter{page}{0}
\begin{flushright}
\hfill DSM-QM434 \\
\hfill DSF/37-98 \\
\hfill October 1998
\end{flushright}
\vspace{.5cm}
\begin{center}{\Large \bf Projective Systems of
Noncommutative Lattices \\ ~ \\
as a Pregeometric Substratum}
\end{center}
\vspace{1cm}
\centerline{\large Giovanni Landi$^{1,3}$, Fedele Lizzi$^{2,3}$}
\vspace{.25cm}
\begin{center}
$^{1}${\it Dipartimento di Scienze Matematiche, Universit\`a di Trieste, \\
P.le Europa 1, I-34127, Trieste, Europe.} \\
landi@univ.trieste.it
\\ ~\\
$^{2}${\it Dipartimento di Scienze Fisiche, Universit\`a di Napoli
Federico II, \\
Mostra d' Oltremare pad. 20, I-80125, Napoli, Europe.} \\
fedele.lizzi@na.infn.it
\\ ~\\
$^{3}${\it INFN, Sezione di Napoli, Napoli, Europe.}
\end{center}
\vspace{1.5cm}
\begin{abstract}
We present an approximation to topological spaces by {\it noncommutative}
lattices. This approximation has a deep physical flavour based on the
impossibility to fully localize particles in any position measurement.
The original space being approximated is recovered out of a projective limit.
\end{abstract}

\vfill
To appear in `{\it Quantum Groups and Fundamental Physical Applications}',
ISI Guccia, Palermo, December 1997, D. Kastler and M. Rosso Eds., 
(Nova Science Publishers, USA).

\newpage
\section{Introduction}
We are going to present a {\it projective paradigm} for a quantum
mechanical scheme of position measurements \cite{ncl}. We take as a fact
that any measurement procedure on a space $M$ can involves only a {\it
finite} number of detectors. Then, with some additional technical
assumption on the nature of the detectors, to any such a system of
detectors, of cardinality $n$, say, we shall associate a topological space
$P_n$ made of $n$ points and endowed with a non {\it trivial} topology.
Here nontriviality mainly means that $P_n$ is not a Hausdorff space (for
the moment we shall not linguistically distinguish between a topology on a
set of point and the set itself) so that it is not possible to isolate
completely its points. Were this the case, and $n$ being finite, one can
only get the trivial topology in which each point is both closed and open
at the same time, so being completely {\it isolated}. In fact, it turn out
that each space is not even $T_1$ but only $T_0$
\footnote{We recall that a topological space is called $T_1$ if any point
of the space is closed. The space is called $T_0$ if for any two distinct
points of it there is an open neighbourhood of one of the points which does
not contain the other.}. On the one hand, the fact that $P_n$ has only a
finite number of points reflects the fact that we get only coarse
information about the space $M$. On the other hand, the nontriviality of
the topology of $P_n$ is reflected in the non vanishing of some of its
homotopy groups which exactly parallel those of $M$. To increase the number
of detectors, so as to get more and more details of the space $M$, results
in the construction of a {\it projective system of topological spaces}. The
limit of the system is a $T_0$ topological space out of which $M$ can be
canonically identified.

Each space $P_n$ being non Hausdorff, there is no room for $\IC$-valued
continuous functions on $P_n$, apart from the constant ones. The crucial
and interesting fact is that there are plenty of operator-valued functions
on $P_n$. Indeed, with any space $P_n$ one associates a noncommutative
$C^*$-algebra $\ca_n$ (In fact, more than one) of operator valued functions
on $P_n$. The space $P_n$ itself can be identified as the space $Prim
\ca_n$ of primitive ideals of $\ca_n$ endowed with the Jacobson topology,
an ideal being called primitive if it is the kernel of an irreducible
representation. Thus, each space $P_n$ is a truly noncommutative space, and
we shall call it a {\it noncommutative lattice} (though this is a bit of a
misnomer because in reality they are not lattices in the usual sense). As
we shall see, the algebras $\ca_n$'s are approximately finite dimensional
(AF) algebras, that is they can be approximated in norm by direct sums of
matrix algebras. This fact allows some sort of a {\it second order
approximation} in so that one can use matrix approximations to do
calculations.

Contrary to what happens in general for noncommutative spaces, which are
{\it characterized by the effective indiscernibility of their elements}
\cite[page 74]{Co1}, in a noncommutative lattice it is possible to discern
its elements (they are indeed finite in number!) and this makes them easy to
`visualize'. However, in a noncommutative lattice there are region of
nonseparability: there are points that cannot be separated from others.

Finally, we cannot resist to quote from \cite{daniel}: {\it I heard Alain
Connes say that he gets a deep hint from algebraic $K$-theory that the
ultimate non-commutative algebra might be of the nature of the discrete
$C^*$-algebras considered by logician}. We believe that the algebras that
we present in this paper are exactly of such a nature.

\section{The Topological Approximation}\label{se:toap}

The idea of a `discrete substratum' underpinning  the `continuum' is
somewhat spread among physicists. With particular emphasis this idea has
been pushed by R. Sorkin who, in \cite{So}, assumes that the substratum be
a {\it finitary} (see later) topological space which maintains some of the
topological information of the continuum. It turns out that the finitary
topology can be equivalently described in terms of a partial order. This
partial order has been alternatively interpreted as determining the causal
structure in the approach to quantum gravity of \cite{BLMS}. Recently,
finitary topological spaces have been interpreted as noncommutative
lattices and noncommutative geometry has been used to construct quantum
mechanical and lattice field theory models, on them \cite{ncl,bbllt}.

Given a suitable covering of a topological space $M$, by identifying any
two points of $M$ which cannot be `distinguished' by the sets in the
covering, one constructs a lattice with a finite (or in general  a
countable) number  of points. Such a lattice, with the quotient topology,
becomes a $T_0$-space which turns out to be the structure space (or
equivalently, the space of primitive ideals) of a postliminal \footnote{It
is a general fact that for a postliminal algebra, irreducible
representations are completely characterized by their kernels \cite{FD}.}
approximately finite dimensional (AF) algebra. Therefore, the lattice is
truly a noncommutative space.

We will have as starting point the fact that it effectively impossible to
localize (at a geometric point) the position of a particle. Detectors in
actual physical situation have always a finite range.
Let us suppose we are about to measure the position of a particle
which moves on a circle, of radius one say, $S^1 = \{ 0\leq\varphi\leq
2\pi, ~{\rm mod}~ 2\pi\}$. Our {\it `detectors'} will be taken to be
(possibly overlapping) open subsets of $S^1$ with some mechanism which
switches on the detector when the particle is in the corresponding open
set. The number of detectors must be clearly finite and, as an example, we
take them to consist of the following three open subsets whose union
covers $S^1$,
\be
U_1 = \{ - {1 \over 3} \pi < \varphi < {2 \over 3} \pi \}, ~~~
U_2 = \{{1 \over 3} \pi < \varphi < {4 \over 3} \pi \},  ~~~
U_3 = \{ \pi < \varphi < 2\pi \}.
\label{detectors}
\ee
Now, if two detectors, $U_1$ and $U_2$ say, are on, we will know that the particle
is in the intersection $U_1 \cap U_2$  although we will be unable to distinguish any
two points in this intersection. The same will be true for the other two intersections.
Furthermore, if only one detector, $U_1$ say, is on, we can infer the presence of the
particle in the {\it closed } subset of $S^1$ given by $U_1 \setminus \{U_1 \cap U_2
\bigcup U_1 \cap U_3 \}$ but again we will be unable to distinguish any
two points in this closed set. The same will be true for the other two closed sets of
similar type. Summing up, if we have only the three detectors (\ref{detectors}), we
are forced to identify the points which cannot be distinguished and $S^1$ will be
represented by a collection of six points $P = \{\alpha, \beta, \gamma, a, b, c \}$
which correspond to the following identifications
\bea
U_1 \cap U_3 = \{{5 \over 3} \pi < \varphi < 2\pi \} & \ra &  \alpha ,
\nonumber\\
U_1 \cap U_2 = \{ {1 \over 3} \pi < \varphi < {2 \over 3} \pi \} & \ra & \beta , \\
U_2 \cap U_3 = \{ \pi < \varphi < {4 \over 3} \pi \} & \ra & \gamma ,
\nonumber\\
U_1 \setminus \{U_1 \cap U_2 \bigcup U_1 \cap U_3 \} = \{ 0 \leq \varphi
\leq {1 \over 3} \pi \} & \ra & a , \nonumber\\
U_2 \setminus \{U_2 \cap U_1 \bigcup U_2 \cap U_3 \} =
\{ {2 \over 3} \pi \leq \varphi \leq \pi \} & \ra & b , \nonumber\\
U_3 \setminus \{U_3 \cap U_2 \bigcup U_3 \cap U_1 \} =
\{ {4 \over 3} \pi \leq \varphi \leq {5 \over 3} \pi \} & \ra & c . \label{discre}
\eea
We can push things a bit further and keep track of the kind of set from which a
point in $P$ comes by declaring the point to be open (respectively closed) if the
subset of $S^1$ from  which it comes is open (respectively closed).
Thus we endow the space $P$ with a topology a basis of
which consists by the following open (by definition)  sets,
\be\label{top6}
\{\alpha \}, ~~\{\beta \}, ~~\{\gamma \}, ~~~
\{\alpha, a, \beta \}, ~~\{\beta, b, \gamma\}, ~~\{\alpha, c, \gamma \}~.
\ee
The corresponding topology on the quotient space $P$ is the quotient
topology of the one on $S^1$ generated by the three open sets
$\{U_1, U_2, U_3\}$, by the
quotient map (\ref{discre}).

In general, let us suppose that we have a topological space $M$ together with
an open covering $\cu =\{U_\lambda\}$ which is also a topology for $M$,
so that $\cu$ is
closed under arbitrary unions and finite intersections.
We define  an equivalence relation among points of $M$ by declaring that any two
points $x, y \in M$ are equivalent if every open set $U_\lambda$
containing either $x$ or $y$ contains the other too,
\be
   x\sim y ~~~~{\rm if~ and~ only~ if}~~~~ x\in U_\lambda \iff y\in U_\lambda~,
~~~\forall~~ U_\lambda \in \cu~ . \label{2.2}
\ee
Thus, two points of $M$ are identified if they cannot be distinguished by
any `detector' in the collection $\cu$. The space $P_{\cu}(M) =: M
/\!\!\sim$ of equivalence classes is then given the quotient topology. If
$\pi : M \ra P_{\cu}(M)$ is the natural projection, a set $U \subset
P_{\cu}(M)$ is declared to be open if and only if $\pi^{-1}(U)$ is  open in
the topology of $M$ given by $\cu$. The quotient topology is the  finest
one making $\pi$ continuous. When $M$ is compact, the covering $\cu$ can be
taken to be finite so that $P_{\cu}(M)$ will consist of a finite number of
points. If $M$ is only locally compact the covering can be taken to be
locally finite and each point has a neighbourhood intersected by only
finitely many $U_\lambda$' s. Then the space $P_{\cu}(M)$ will consist of a
countable number of points; in the terminology of \cite{So} $P_{\cu}(M)$
would be a {\it finitary}
 approximation of $M$. If
$P_{\cu}(M)$ has
$N$ points we shall also denote it by $P_N(M)$ although this notation is
incomplete since it does not keep track of the topology given on the set of $N$
points. For the examples considered in these paper, the topology
will always be given explicitly.
For example, the finite space given by (\ref{discre}) is $P_6(S^1)$.

In general, $P_{\cu}(M)$ is not Hausdorff: from (\ref{top6}) it is evident
that in $P_6(S^1)$, for instance, we cannot isolate the point $a$ from $\alpha$ by
using open sets. It is not even a $T_1$-space; again, in $P_6(S^1)$ only the points
$a$, $b$ and $c$ are closed while the points $\alpha$, $\beta$ and $\gamma$ are open. In
general there will be points which are neither closed nor open.
However, $P_{\cu}(M)$ is always a $T_0$-space, being, indeed, the $T_0$-quotient
of $M$ with respect to the topology $\cu$ \cite{So}.

\section{Order and Topology}\label{se:orto}
What we shall show next is how the topology of any finitary
$T_0$ topological space $P$ can be given equivalently by means of a partial order
which makes $P$ a
{\it partially ordered set} (or {\it poset} for short).
Consider first the case when $P$ is finite. Then, the collection $\tau$ of open sets (the
topology on $P$) will be closed under arbitrary unions and arbitrary intersections.
Thus, for any point $x\in P$, the intersection of all open sets
containing it,
\be\label{basis}
\Lambda(x) =: \bigcap\{U \in \tau ~|~ x \in U \}~,
\ee
will be the smallest open set containing the point.
A relation $\preceq$ is defined on $P$ by
\be\label{order1}
x \preceq y~ \Leftrightarrow ~\Lambda(x) \subseteq \Lambda(y)~, ~~\forall ~x,y \in P~.
\ee
Now, $x\in \Lambda(x)$ always, so that the previous definition is equivalent to
\be\label{order2}
x \preceq y~ \Leftrightarrow ~x \in \Lambda(y)~,
\ee
which can also be stated saying that
\be\label{order3}
x \preceq y  ~ \Leftrightarrow ~ {\rm every ~open ~set
~containing} ~y~ {\rm also ~contains} ~x~,
\ee
or, in turn, that
\be\label{order4}
x \preceq y  ~ \Leftrightarrow ~ y \in \bar{\{ x\}}~,
\ee
with $\bar{\{ x\}}$ the closure of the one point set $\{ x\}$. Another
equivalent definition can be given by saying that $x \preceq y$ if and only if the
constant sequence
$(x, x, x, \cdots)$ converges to $y$. It is worth noticing that in a $T_0$-space the
limit of a sequence need not be unique so that the constant sequence $(x, x, x,
\cdots)$ may converge to more than one point.
~\\
>From (\ref{order1}) it is clear that the relation $\preceq$ is reflexive,
and transitive.
Furthermore, since $P$ is a $T_0$-space, for any two distinct points $x,y\in P$, there
is at least one open set containing $x$, say, and not $y$. This, together with
(\ref{order3}), implies that the relation $\preceq$ is symmetric as well,
$x \preceq y~, ~y \preceq x ~\Rightarrow~ x = y$.
Summing up, we see that a $T_0$ topology on a finite space $P$ determines a reflexive,
antisymmetric and transitive relation, namely a
{\it partial order}.
Conversely, given a partial order $\preceq$ on the set $P$, one produces a
topology on $P$ by taking as a basis for it the finite collection of `open'
sets defined by
\be\label{bo}
\Lambda(x) =: \{y \in P ~|~ y \preceq x \}~, ~~ \forall ~x \in P~.
\ee
Thus, a subset $W\subset P$ will be open if and only if it is the union of sets of the
form (\ref{bo}), that is, if and only if $x\in W$ and $y\preceq x ~\Rightarrow~ y
\in W$.
Indeed, the smallest open set containing $W$ is given by
$\Lambda(W) = \bigcup_{x\in W} \Lambda(x)$,
and $W$ is open if and only if $W = \Lambda(W)$.\\
The resulting topological space is clearly $T_0$ by the antisymmetry of the order
relation.

It is easy to express the closure operation in terms of the partial order. From
(\ref{order4}), the closure $V(x) = \bar{\{ x\}}$, of the one point set $\{ x\}$ is
given by
\be\label{smcl}
V(x) =: \{y \in P ~|~ x \preceq y \}~, ~~ \forall ~x \in P~.
\ee
A subset $W \subset P$ will be closed if and only if  $x\in W$ and $x\preceq y
~\Rightarrow~ y \in W$. Indeed, the closure of $W$ is given by
$ V(W) = \bigcup_{x \in W} V(x)$,
and $W$ is closed if and only if $W = V(W)$.

If one relaxes the condition of finiteness of the space $P$, there is still
an equivalence between topology and partial order for any $T_0$ topological
space which has the additional property that every intersection of open
sets is an open set (or equivalently, that every union of closed sets is a
closed set), so that the sets (\ref{basis}) are all open and provide a
basis for the topology \cite{Al,Bo1}. This would be the case if $P$ were a
finitary approximation of a (locally compact) topological space $M$,
obtained then from a locally finite covering of $M$.

A pictorial representation of the topology of a poset is obtained by constructing the
associated
{\it Hasse diagram}: one  arranges the points of the
poset at different levels and connects them by the rules :
($x\prec y$ will indicate that $x$ precedes $y$ while $x\not=y$)
\begin{enumerate}
\item
if $x\prec y$, then $x$ is at a lower level than $y$;
\item
if $x\prec y$ and there is no $z$ such that $x\prec z\prec y$,
then $x$ is at the level immediately below $y$ and these two points
are connected by a link.
\end{enumerate}
\begin{figure}[t]
\begin{picture}(300,110)(-260,-55)
\put(-30,30){\circle*{4}}
\put(30,30){\circle*{4}}
\put(-30,-30){\circle*{4}}
\put(30,-30){\circle*{4}}
\put(-30,30){\line(0,-1){60}}
\put(30,30){\line(0,-1){60}}
\put(-30,30){\line(1,-1){60}}
\put(30,30){\line(-1,-1){60}}
\put(-30,35){$x_3$}
\put(30,35){$x_4$}
\put(-30,-45){$x_1$}
\put(30,-45){$x_2$}
\put(-250,30){\circle*{4}}
\put(-190,30){\circle*{4}}
\put(-250,-30){\circle*{4}}
\put(-190,-30){\circle*{4}}
\put(-130,30){\circle*{4}}
\put(-130,-30){\circle*{4}}
\put(-250,30){\line(0,-1){60}}
\put(-190,30){\line(0,-1){60}}
\put(-130,30){\line(0,-1){60}}
\put(-250,30){\line(1,-1){60}}
\put(-190,30){\line(1,-1){60}}
\put(-250,-30){\line(2,1){120}}
\put(-250,35){$a$}
\put(-250,-45){$\alpha$}
\put(-190,35){$b$}
\put(-190,-45){$\beta$}
\put(-130,35){$c$}
\put(-130,-45){$\gamma$}
\end{picture}
\caption{The Hasse diagrams for $P_6(S^1)$  and for $P_4(S^1)$ }
\label{fi:cirhas}
\end{figure}
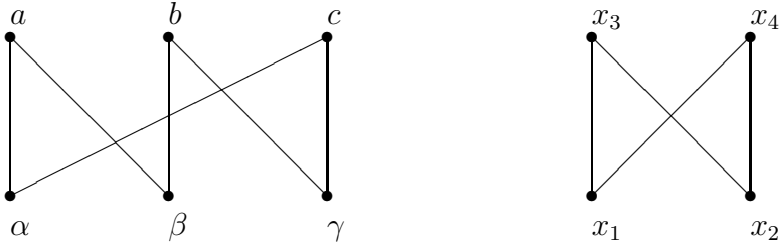
Figure~\ref{fi:cirhas} shows the Hasse diagram for $P_6(S^1)$ whose basis of open
sets is in (\ref{top6}) and for $P_4(S^1)$. For the former, the partial order reads
$\alpha \prec a, ~\alpha \prec c, ~\beta \prec a, ~\beta \prec b, ~\gamma \prec b,
~\gamma
\prec c$. The latter is a four point  approximation of $S^1$ obtained from a covering
consisting of two intersecting open sets. The partial order reads $x_1\prec x_3,
~x_1\prec x_4, ~x_2\prec x_3, ~x_2\prec x_4~$.\\
In Fig.~\ref{fi:cirhas}, (and in general, in any
Hasse diagram the smallest open set
containing any point $x$ consists of all points which are below the given one, $x$, and
can be connected to it by a series of links. For example, for $P_4(S^1)$, we have
as the minimal open sets,
\be\label{top4}
\Lambda(x_1) = \{x_1\}~, ~~~ \Lambda(x_2) = \{x_2\}~, ~~~
\Lambda(x_3) = \{x_1,x_2,x_3\}~, ~~~ \Lambda(x_4) = \{x_1,x_2,x_4\}~,
\ee
which are a basis for the topology of $P_4(S^1)$.

\begin{figure}[th]
\begin{picture}(340,190)(-160,-20)
\put(-88,150){$U_{i-1}$}
\put(-32,115){$U_{i}$}
\put(28,150){$U_{i+1}$}
\put(88,115){$U_{i+2}$}
\put(-160,135){$\dots$}
\put(-140,135){\line(1,0){280}}
\put(150,135){$\dots$}
\put(-130,135){$($}
\put(-110,126){{\Large$)$}}
\put(-70,126){{\Large$($}}
\put(-50,135){$)$}
\put(-10,135){$($}
\put(10,126){{\Large$)$}}
\put(50,125){{\Large$($}}
\put(70,135){$)$}
\put(110,135){$($}
\put(130,126){{\Large$)$}}
\put(10,80){$\pi$}
\put(0,100){\vector(0,-1){40}}
\put(-120,0){\circle*{4}}
\put(-60,0){\circle*{4}}
\put(0,0){\circle*{4}}
\put(60,0){\circle*{4}}
\put(120,0){\circle*{4}}
\put(-90,30){\circle*{4}}
\put(-30,30){\circle*{4}}
\put(30,30){\circle*{4}}
\put(90,30){\circle*{4}}
\put(-160,10){$\cdots$}
\put(150,10){$\cdots$}
\put(-90,30){\line(1,-1){30}}
\put(-90,30){\line(-1,-1){30}}
\put(-30,30){\line(1,-1){30}}
\put(-30,30){\line(-1,-1){30}}
\put(30,30){\line(1,-1){30}}
\put(30,30){\line(-1,-1){30}}
\put(90,30){\line(1,-1){30}}
\put(90,30){\line(-1,-1){30}}
\put(-120,0){\line(-1,1){20}}
\put(120,0){\line(1,1){20}}
\put(-120,-10){$x_{i-2}$}
\put(-60,-10){$x_{i-1}$}
\put(0,-10){$x_i$}
\put(60,-10){$x_{i+1}$}
\put(120,-10){$x_{i+2}$}
\put(-90,37){$y_{i-2}$}
\put(-30,37){$y_{i-1}$}
\put(30,37){$y_i$}
\put(90,37){$y_{i+1}$}
\end{picture}
\caption{The finitary poset of the line $\IR$ }
\label{fi:linhas}
\end{figure}
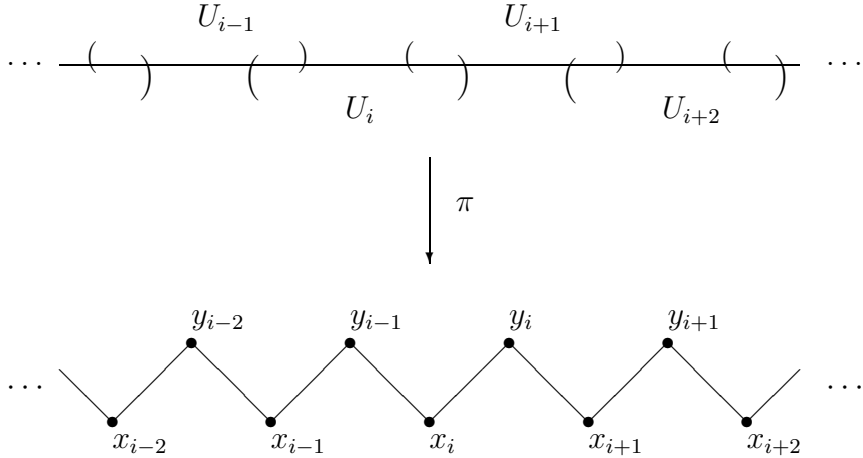

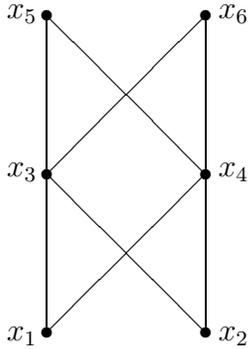
\begin{figure}[t]
\begin{picture}(100,152)(-55,-43)
\put(-30,90){\circle*{4}}
\put(30,90){\circle*{4}}
\put(-30,30){\circle*{4}}
\put(30,30){\circle*{4}}
\put(-30,-30){\circle*{4}}
\put(30,-30){\circle*{4}}
\put(-30,90){\line(0,-1){60}}
\put(30,90){\line(0,-1){60}}
\put(-30,90){\line(1,-1){60}}
\put(30,90){\line(-1,-1){60}}
\put(-30,30){\line(0,-1){60}}
\put(30,30){\line(0,-1){60}}
\put(-30,30){\line(1,-1){60}}
\put(30,30){\line(-1,-1){60}}
\put(-45,89){$x_5$}
\put(35,89){$x_6$}
\put(-45,29){$x_3$}
\put(35,29){$x_4$}
\put(-45,-33){$x_1$}
\put(35,-33){$x_2$}
\end{picture}
\caption{The Hasse diagram for the poset $P_6(S^2)$ }
\label{fi:sphpos}
\end{figure}

\noindent
The generic finitary poset $P(\IR)$ associated with the real line $\IR$ is shown in
Fig.~\ref{fi:linhas}. The corresponding projection $\pi : \IR \ra P(\IR)$ is given by
\be
\begin{array}{rcl}
U_i \cap U_{i+1} & \lra &  x_i ~,~~i \in \IZ~,  \\
U_{i+1}\setminus\{ U_i \cap U_{i+1}  \bigcup U_{i+1} \cap U_{i+2} \}
& \lra &  y_i ~,~~i \in \IZ~.
\end{array}
\ee
A basis for the quotient topology is provided by the collection of all open sets of the
form
\be
\Lambda(x_i) = \{x_i \}~, ~~\Lambda(y_i) = \{x_{i}, y_i, x_{i+1} \} ~,~~i \in \IZ~.
\ee
\noindent
Figure~\ref{fi:sphpos} shows  the Hasse diagram for the six-point poset
$P_6(S^2)$ of the two dimensional sphere, coming from a covering with four open
sets, which was derived in \cite{So}. A basis for its topology is given by
\bea
&&\Lambda(x_1) = \{x_1\}~, ~~\Lambda(x_2) = \{x_2\}~,
~~\Lambda(x_3) = \{x_1,x_2,x_3\}~, ~~\Lambda(x_4) = \{x_1,x_2,x_4\}~, \nonumber \\
&&~\nonumber \\
&& \Lambda(x_5) = \{x_1,x_2,x_3,x_4,x_5\}~, ~~\Lambda(x_6) = \{x_1,x_2,x_3,x_4,x_6\}~.
\label{2.6}
\eea
The top two points are closed, the bottom two points are open and the
intermediate ones are neither closed nor open.

One key feature of noncommutative lattices is that, although composed of a
finite number of elements, not all of the topological information of the
original set has disappeared. For example, one can prove that for the first
homotopy group, $\pi_1(P_N(S^1)) = \IZ = \pi(S^1)$ whenever $N \geq 4$
\cite{So}.

\section{The Reconstruction of the Approximated Space}\label{se:recospa}
We shall now briefly describe how the topological space being approximated
can be recovered `in the limit' by considering a sequence of finer and finer
coverings,  the appropriate framework being that of projective (or inverse)
systems of topological spaces \cite{So}.

Let us consider a topological space $M$ and
a sequence $\{ \cu_n \}_{n \in \IN}$ of finer and finer coverings, that is
of coverings such that
\be
\cu_i \subseteq \tau(\cu_{i+1})~,
\ee
where $\tau(\cu)$ is the topology generated by the covering $\cu$.
Here we are relaxing the harmless assumption made in Sect.~\ref{se:toap} that each
$\cu$ is already a subtopology, namely that $\cu = \tau(\cu)$.

\begin{figure}[t]
\begin{picture}(310,310)(-130,-100)
\put(-120,180){$M$}
\put(170,180){$P_\infty$}
\put(-60,180){\vector(1,0){180}}
\put(-100,165){\vector(1,-1){120}}
\put(-100,165){\vector(1,-2){120}}
\put(25,30){$P_j$}
\put(25,-90){$P_i$}
\put(30,130){$\vdots$}
\put(30,110){\vector(0,-1){60}}
\put(30,10){\vector(0,-1){60}}
\put(25,190){$\pi_{\infty}$}
\put(35,-20){$\pi_{ij}$}
\put(-60,30){$\pi_{i}$}
\put(-30,70){$\pi_{j}$}
\put(105,30){$\pi_{i\infty}$}
\put(80,70){$\pi_{j\infty}$}
\put(160,165){\vector(-1,-1){120}}
\put(160,165){\vector(-1,-2){120}}
\end{picture}
\caption{The projective system of topological spaces with continuous maps which
approximates the space $M$}
\label{fi:invlim}
\end{figure}
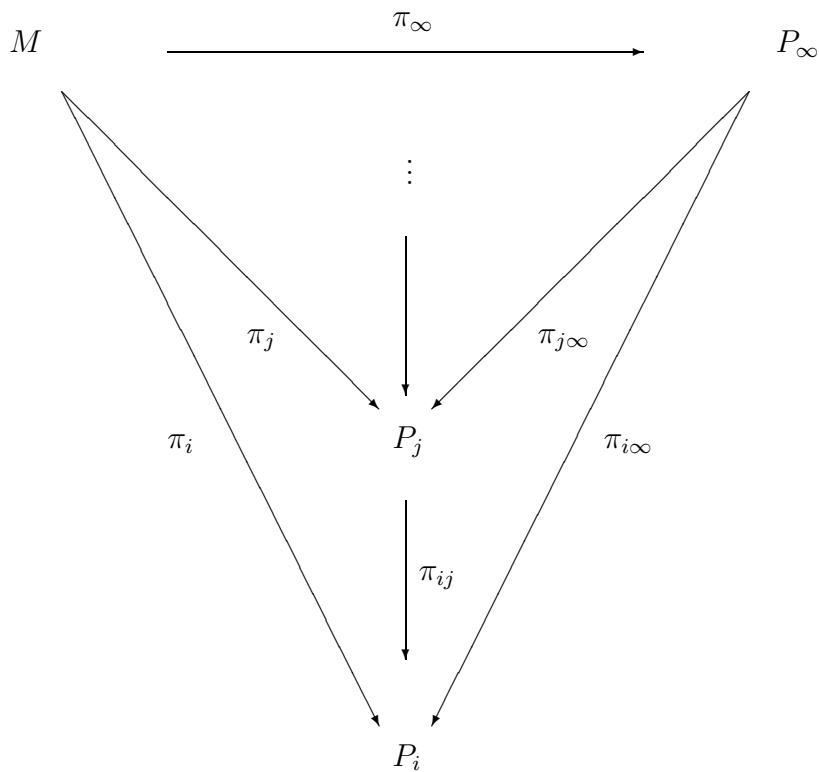

In Sect.~\ref{se:toap} we have associated with each covering $\cu_i$ a
$T_0$-topological space $P_i$ and a continuous surjection
\be
\pi_i : M \ra P_i~.
\ee
We now construct a {\it projective system of spaces $P_i$ together with continuous
maps}
\be
\pi_{ij} : P_j \ra P_i~,
\ee defined whenever $i \leq j$ and such that
\be
\pi_i = \pi_{ij} \circ \pi_j~. \label{prmap}
\ee
These maps are uniquely defined by the fact that the spaces $P_i$'s are $T_0$ and
that the map $\pi_i$ is continuous with respect to $\tau(\cu_j)$ whenever $i\leq j$.
Indeed, if $U$ is open in $P_i$, then $\pi_i^{(-1)}(U)$ is open in the
$\cu_i$-topology by definition, thus it is also open in the finer $\cu_j$-topology
and $\pi_i$ is continuous in $\tau(\cu_j)$.  Furthermore, uniqueness also
implies the compatibility conditions
\be
\pi_{ij} \circ \pi_{jk} = \pi_{ik}~,
\ee
whenever $i \leq j \leq k$. Indeed, the map $\pi_{ij}$ is the solution (by
definition it is then unique) of a universal mapping problem for maps relating
$T_0$-spaces \cite{So}.
>From the surjectivity of the maps $\pi_i$'s and the relation
(\ref{prmap}), it follows that all maps $\pi_{ij}$ are surjective. \\
The projective system of topological spaces together with continuous maps
$\{P_i, \pi_{ij} \}_{i,j \in \IN}$ has a unique
{\it projective limit}, i.e. a
topological space $P_\infty$, together with continuous maps
\be
\pi_{i\infty} : P_\infty \ra P_i~,
\ee
such that
\be
\pi_{ij} \circ \pi_{j\infty} = \pi_{i\infty}~,
\ee
whenever $i \leq j$. The space $P_\infty$ and the maps $\pi_{ij}$ can be
constructed explicitly. An element $x \in P_\infty$ is an arbitrary coherent sequence
of elements
$x_i  \in P_i$,
\be
x = (x_i)_{i\in\IN} ~,~ x_i \in P_i ~|~ \exists ~N_0 ~~~{\rm s.t.}~~~ x_i =
\pi_{i,i+1}(x_{i+1})~, ~~\forall ~ i \geq N_0~.
\ee
As for the map $\pi_{i\infty}$, it  is simply defined by
\be
\pi_{i\infty}(x) = x_i~.
\ee
The space $P_{\infty}$ is made into a $T_0$ topological space by endowing it with the
weakest topology making all maps $\pi_{i\infty}$ continuous: a basis for it is given
by the sets $\pi^{(-1)}_{i\infty}(U)$, for all open sets $U \subset P_i$.
The projective system and its limit are depicted in Fig.~\ref{fi:invlim}.
\\
It turns out that the limit space $P_{\infty}$ is {\it bigger} than the starting
space $M$ and that the latter is contained as a dense subspace. Furthermore, $M$ can be
characterized as the set of all {\it closed} points of $P_{i\infty}$.
First of all, we also get a unique (by universality) continuous map
\be
\pi_\infty : M \ra P_\infty~,
\ee
which satisfies
\be
\pi_i = \pi_{i\infty} \circ \pi_\infty~, ~~~ \forall ~ i \in \IN~.
\ee
The map $\pi_\infty$ is the `limit' of the maps $\pi_i$. However, while the latter are
surjective, under mild hypothesis, the former turns out to be {\it injective}.
We have, indeed, the following results whose proof is in
\cite{So,gianni}.
\bprop
The image $\pi_\infty(M)$ is dense in $P_\infty$.
\eprop
\bprop\label{lemma2}
Let $M$ be $T_0$ and the collection $\{\cu_i \}$ of coverings be such that for every $m
\in M$ and every neighbourhood $N \ni m$, there exists an index $i$ and an element $U
\in \tau(\cu_i)$ such that $m \in U \subset N$. Then, the map $\pi_\infty$ is injective.
\eprop

\noindent
In a sense, the second condition in the previous Proposition just says
that the covering $\cu_i$ contains `enough small open sets', a condition one would
expect in the process of recovering $M$ by a refinement of the coverings.

As alluded to before, there is a nice characterization of the points of $M$ (or better
still of $\pi_\infty(M)$) as the set of all closed points of $P_\infty$.
\bprop
Let $M$ be $T_1$ and let the collection $\{\cu_i \}$ of coverings fulfil
the `fineness' condition of Proposition~\ref{lemma2}. Let each covering
$\cu_i$ consist only of sets which are bounded (have compact closure). Then
$\pi_\infty : M \ra P_\infty$ embeds $M$ in $P_\infty$ as the subspace of
closed points.
\eprop

As for the extra points of $P_\infty$, one can prove that for any extra
$y \in P_\infty$, there exists an $x\in\pi_\infty(M)$ to which $y$ is `infinitely
close'. Indeed, $P_\infty$ can be turned into a poset by defining a partial order
relation as follows
\be
x \preceq_\infty y ~~~\Leftrightarrow~~~ x_i \preceq y_i~, ~~~\forall ~i~,
\label{poinf}
\ee
where the coherent sequences $x=(x_i)$ and $y=(y_i)$ are any two elements of
$P_\infty$. In fact, one could directly construct $P_\infty$ as the
projective limit of a
projective system of posets by defining a partial order on
the coherent sequences as in
(\ref{poinf}).

Then one can characterize $\pi_\infty(M)$ as the {\it set of maximal elements of
$P_\infty$}, with respect to the order $\preceq_\infty$. Given any such maximal
element $x$, the points of $P_\infty$ which are infinitely close to $x$ are all (non
maximal) points which converge to $x$, namely all (not maximal) $y\in P_\infty$ such
that $y \preceq_\infty x$. In $P_\infty$, these points $y$ cannot be separated from
the corresponding $x$. By identifying points in $P_\infty$ which cannot be separated
one recovers $M$.  The interpretation that emerges is that the top points of a poset
$P(M)$ (which are always closed) approximate the points of $M$ and give all of $M$ in
the limit. The r\^ole of the remaining points is to `glue' the top points together so
as to produce a topologically nontrivial approximation to $M$. They also give the extra
points in the limit.

In \cite{pangs} a somewhat different interpretation of the approximation and of the
limiting procedure in terms of simplicial decompositions has been proposed.

\section{Noncommutative Lattices}\label{se:ncl1}

It turns out that any (finite) poset $P$ is the structure space
$\ha$ (the space of irreducible representations) of a
noncommutative
$C^*$-algebra $\ca$ of operator valued functions which then plays the r\^ole of the
algebra of continuous functions on $P$. It is worth noticing that, a poset $P$
being non Hausdorff, there cannot be `enough'
$\IC$-valued continuous functions on $P$ since the latter separate points. For
instance, on the poset of Fig.~\ref{fi:cirhas} or Fig.~\ref{fi:sphpos} the only
$\IC$-valued continuous functions are the constant ones. In fact, the previous
statement is true for each connected component of any poset.

Indeed, there is a complete classification of all separable
$C^*$-algebras with a finite dual \cite{BL}. Given
any finite $T_0$-space $P$, it is possible to construct a $C^*$-algebra $\ca(P, d)$ of
operators on a separable
Hilbert space $\ch(P, d)$ which satisfies
$\widehat{\ca(P, d)} = P$. Here $d$ is a function on $P$ with values in $\IN \cup \infty$
which is called a {\it defector}. Thus there is more than one algebra with the same
structure space. We refer to \cite{BL,ELTfunctions} for the actual
construction of the algebras together with extensions to countable posets. Here, we
shall instead describe a more general class of algebras, namely approximately finite
dimensional ones, a subclass of which is associated with posets. As the name suggests,
these algebras can be approximated by finite dimensional algebras, a fact which has
been used in the construction of physical models on posets as we shall describe in
Sect.~\ref{se:qmm}. %

Before we proceed, we mention that if a separable $C^*$-algebra has a finite dual than
it is postliminal \cite{BL}. As alluded to already, for any such algebra
$\ca$, irreducible representations are completely characterized
by their kernels so that
the space of irreducible representations
is homeomorphic with the space $\prim$
of primitive ideals. Furthermore, the Jacobson topology on $\prim$ is equivalent to the partial
order defined by the inclusion of ideals. This fact in a sense `closes the circle'
making any poset, when thought of as $\prim$ of a noncommutative algebra
$\ca$, a
truly noncommutative space or, rather, a
{\it  noncommutative lattice}.

   \subsection{AF-Algebras}\label{se:afa}
In this Section we shall describe approximately finite dimensional algebras following
\cite{Br1}. A general algebra of this sort may have a rather complicated ideal structure
and a complicated primitive ideal structure. As mentioned before, for applications
to posets only a special subclass is selected.
\bdefi
A $C^*$-algebra $\ca$ is said to be {\it approximately finite dimensional}
(AF) if there exists an increasing sequence
\be
\ca_0 ~{\buildrel I_0 \over \hookrightarrow}~ \ca_1
      ~{\buildrel I_1 \over \hookrightarrow}~ \ca_2
      ~{\buildrel I_2 \over \hookrightarrow}~ \cdots
      ~{\buildrel I_{n-1} \over \hookrightarrow}~ \ca_n
      ~{\buildrel I_n \over \hookrightarrow} \cdots
\label{af}
\ee
of finite dimensional $C^*$-subalgebras of $\ca$, such that $\ca$ is the norm closure
of $\bigcup_n \ca_n~, ~ \ca = \bar{\bigcup_n \ca_n}$. The maps $I_n$ are injective
$^*$-morphisms.
\edefi

\noindent
The algebra $\ca$ is the {\it inductive} (or {\it direct}) {\it limit} of
the {\it inductive} {\it system} $\{\ca_n, I_n \}_{n\in
\IN}$ of algebras \cite{W-O}. As a set, $\bigcup_n \ca_n$ is made of
coherent sequences,
\be
\bigcup_n \ca_n = \{ a=(a_n)_{n \in \IN}~, a_n \in \ca_n ~|~ \exists  N_0
~, a_{n+1} =  I_n(a_n)~, \forall ~n>N_0 \}.
\ee
Now the sequence $(||a_n||_{\ca_n})_{n \in \IN}$ is
eventually decreasing since $||a_{n+1}|| \leq ||a_n||$ (the maps $I_n$ are
norm decreasing) and therefore convergent. One writes for the norm on $\ca$,
\be
||(a_n)_{n \in \IN}|| = \lim_{n \ra \infty} ||a_n||_{\ca_n}~. \label{norm}
\ee
Since the maps $I_n$ are injective, the expression (\ref{norm}) gives a true
norm directly and not simply a seminorm and there is no need to quotient out the zero
norm elements.

\noindent
We shall assume that the algebra $\ca$ has a unit $\II$. If $\ca$ and $\ca_n$ are as
before, then $\ca_n + \IC \II$ is clearly a finite dimensional $C^*$-subalgebra
of $\ca$ and moreover, $\ca_n \subset \ca_n + \IC \II \subset \ca_{n+1} + \IC \II$. We
may thus assume that each $\ca_n$  contains the unit $\II$ and that the maps
$I_n$ are unital.
\bexam
Let $\ch$ be an infinite dimensional (separable) Hilbert space. The algebra
\be\label{alge}
\ca = \ck(\ch) + \IC \II_\ch~,
\ee
with $\ck(\ch)$ the algebra of compact operators, is an AF-algebra \cite{Br1}. The
approximating algebras are given by
\be
\ca_n = \IM_{n}(\IC) \oplus \IC~, ~~n > 0~,
\ee
with embedding
\be\label{dentro}
\IM_{n}(\IC) \oplus \IC \ni (\Lambda, \lambda) \mapsto
\left( \left\{
\begin{array}{ll}
\Lambda & 0 \\
0 & \lambda
\end{array}
\right\}, \lambda
\right)
\in \IM_{n+1}(\IC) \oplus \IC~.
\ee
Indeed, let $\{\xi_n\}_{n\in\IN}$ be an orthonormal basis in $\ch$ and let $\ch_n$ be
the subspace generated by the first $n$ basis elements, $\{\xi_1, \cdots, \xi_n\}$.
With $\cp_n$ the orthogonal projection onto $\ch_n$, define
\be
\begin{array}{lcl}
\ca_n & = &
\{T \in \cb(\ch) ~|~ T(\II - \cp_n) = (\II - \cp_n)T \in \IC (\II - \cp_n) \} \\
~     & \simeq &  \cb(\ch_n) \oplus \IC \simeq \IM_{n}(\IC) \oplus \IC~.
\end{array}
\ee
Then $\ca_n$ embeds in $\ca_{n+1}$ as in (\ref{dentro}). Since each $T\in\ca_n$ is a
sum of a finite rank operator and a multiple of the identity, one has that
$\ca_n \subseteq \ca = \ck(\ch) + \IC\II_\ch$ and, in turn,
$\bar{\bigcup_n \ca_n} \subseteq \ca = \ck(\ch) + \IC\II_\ch$. Conversely, since finite
rank operators are norm dense in $\ck(\ch)$, and finite linear combinations of strings
$\{\xi_1, \cdots, \xi_n \}$ are dense in $\ch$, one gets that
$\ck(\ch) + \IC\II_\ch \subset \bar{\bigcup_n \ca_n}$.

\begin{figure}[t]
\begin{picture}(55,93)(-10,-43)
\put(30,30){\circle*{4}}
\put(0,-30){\circle*{4}}
\put(30,30){\line(-1,-2){30}}
\put(35,29){$\ci_2$}
\put(5,-33){$\ci_1$}
\end{picture}
\caption{The two point poset of the interval }
\label{fi:point}
\end{figure}
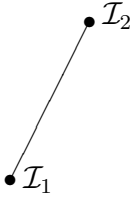

The algebra (\ref{alge}) has only two irreducible representations
\cite{BL},
\be\label{repalge}
\begin{array}{ll}
\pi_1 : \ca \lra \cb(\ch)  ~, & a = (k + \lambda \II_\ch)
\mapsto \pi_1(a) = a~, \\
\pi_2 : \ca \lra \cb(\IC) \simeq \IC ~, & a = (k + \lambda \II_\ch)
\mapsto \pi_2(a) = \lambda~,
\end{array}
\ee
with $\lambda_1,\lambda_2 \in \IC$ and $k \in \ck(\ch)$; the corresponding kernels
being
\be\label{kerint}
\ci_1 =: ker(\pi_1) = \{0\} ~, ~~~
\ci_2 =: ker(\pi_2) = \ck(\ch) ~.
\ee
The partial order given by the inclusions $\ci_1 \subset \ci_2$ produces the two point
poset shown in Fig.~\ref{fi:point}.
As we shall see, this space is really the fundamental building
block for all posets. A comparison with the poset of the line in
Fig.~\ref{fi:linhas}, shows that it can be thought of as a two point approximation of
an interval.
\eexam

In general, each subalgebra $\ca_n$ being a finite dimensional $C^*$-algebra, is a
direct sum of  matrix algebras,
\be
\ca_n = \bigoplus_{k=1}^{k_n} \IM_{d_k^{(n)}}(\IC)~,
\ee
where $\IM_d(\IC)$ is the algebra of $d \times d $ matrices with complex coefficients.
In order to study the embedding $\ca_1 \hookrightarrow \ca_2$ of any two such algebras
$\ca_1 = \bigoplus_{j=1}^{n_1} \IM_{d_j^{(1)}}(\IC)$ and
$\ca_2 = \bigoplus_{k=1}^{n_2} \IM_{d_k^{(2)}}(\IC)$, one uses the fact that it is
always possible \cite{gianni} to choose bases in $\ca_1$ and $\ca_2$ in such a way
as to identify
$\ca_1$ with a subalgebra of
$\ca_2$ having the following form
\be
\ca_1 \simeq \bigoplus_{k=1}^{n_2} \left( \bigoplus_{j=1}^{n_1} N_{kj}
\IM_{d_j^{(1)}}(\IC) \right)  \; .
\ee
Here, with any two nonnegative integers $p,q$, the symbol $p\IM_{q}(\IC)$ stands for
\be\label{azzo}
p\IM_{q}(\IC) \simeq \IM_{q}(\IC) \otc \II_p~,
\ee
and one identifies $\bigoplus_{j=1}^{n_1} N_{kj} \IM_{d_j^{(1)}}(\IC)$ with a
subalgebra of $\IM_{d_k^{(2)}}(\IC)$. The nonnegative integers $N_{kj}$ satisfy the
condition
\be
\sum_{j=1}^{n_1} N_{kj} d^{(1)}_j = d^{(2)}_k  \; . \label{dim}
\ee
One says that the algebra $\IM_{d_j^{(1)}}(\IC)$ is {\it partially
embedded} in $\IM_{d_k^{(2)}}(\IC)$ with {\it multiplicity} $N_{kj}$. A
useful way of representing the algebras $\ca_1$, $\ca_2$ and the embedding
$\ca_1 \hookrightarrow \ca_2$ is by means of a diagram, the so called {\it
Bratteli diagram} \cite{Br1}, which can be constructed out of the
dimensions $d_j^{(1)}~, ~j=1,\ldots,n_1$ and $d_k^{(2)}~, ~k=1,\ldots,n_2$,
of the diagonal blocks of the two algebras and out of the numbers $N_{kj}$
that describe the partial embeddings. One draws two horizontal rows of
vertices, the top (bottom) one representing $\ca_1$ ($\ca_2$) and
consisting of $n_1$ ($n_2$) vertices, one for each block which are labelled
by the corresponding dimensions $d_1^{(1)}, \ldots, d_{n_1}^{(1)}$
($d_1^{(2)},\ldots,d_{n_2}^{(2)}$). Then, for each $j=1,\ldots,n_1$ and
$k=1,\ldots,n_2$, the relation $d^{(1)}_j \searrow^{N_{kj}} d^{(2)}_k$
denotes the embeddings of $\IM_{d^{(1)}_j}(\IC)$ in $\IM_{d^{(2)}_k}(\IC)$
with multiplicity $N_{kj}$.

For any AF-algebra $\ca$ one repeats the procedure for each
level, and in this way one obtains a semi-infinite diagram, denoted by $\cd(\ca)$ which
completely defines $\ca$ up to isomorphism. The diagram
$\cd(\ca)$ depends not only on the
collection of $\ca$'s but also on the particular sequence $\{\ca_n\}_{n \in \IN}$ which
generates $\ca$.
However, one can obtain an algorithm which allows one to construct from
a given diagram all diagrams which define AF-algebras which are isomorphic with the
original one \cite{Br1}. The problem of identifying the limit algebra or of
determining whether or not two such limits are isomorphic can be very subtle. Elliot
\cite{El} has devised an invariant for AF-algebras in terms of the
corresponding $K$ theory which completely distinguishes among them (see also \cite{Ef}).
%
%
It is worth remarking that the
isomorphism class of an AF-algebra $\bar{\bigcup_n \ca_n}$ depends not only on the
collection of algebras $\ca_n$'s but also on the way they are embedded into each other.

Given a set $\cd$ of ordered pairs $(n, k), k = 1, \cdots, k_n~, ~n = 0, 1, \cdots$,
with $k_0=1$, and a sequence $\{ \searrow^p \}_{p = 0, 1, \cdots }$ of relations on
$\cd$, the latter is the diagram $\cd(\ca)$ of an AF-algebras when the following
conditions are satisfied, \label{`brdiagram'}
\begin{description}
\item[(i)] If $(n, k), (m, q) \in \cd$ and $m = n+1$, there exists one and only one
nonnegative (or equivalently, at most a positive) integer $p$ such that
$(n, k) \searrow^p (n+1, q)$.
\item[(ii)] If $m \not= n+1$, no such integer exists.
\item[(iii)] If $(n, k) \in \cd$, there exists $q \in \{1, \cdots, n_{n+1} \}$ and a
nonnegative integer $p$ such that $(n, k) \searrow^p (n+1, q)$.
\item[(iv)] If $(n, k) \in \cd$ and $n > 0$, there exists $q \in \{1, \cdots, n_{n-1}
\}$ and a nonnegative integer $p$ such that $(n-1, q) \searrow^p (n, k) $.
\end{description}
~\\

It is easy to see that the diagram of a given AF-algebra satisfies the previous
conditions. Conversely, if the set $\cd$ of ordered pairs satisfies these properties,
one constructs by induction a sequence of finite dimensional
$C^*$-algebras $\{ \ca_n \}_{n\in\IN}$  and of injective morphisms $I_n : \ca_n \ra
\ca_{n+1}$ in such a manner so that the inductive limit $\{ \ca_n, I_n \}_{n\in\IN}$ will
have $\cd$ as its diagram. Explicitly, one defines
\be
\ca_n = \bigoplus_{k; (n,k) \in \cd} \IM_{d_k^{(n)}}(\IC) = \bigoplus_{k=1}^{k_n}
\IM_{d_k^{(n)}}(\IC)~,
\ee
and morphisms
\be
\begin{array}{l}
I_n : \bigoplus_{j=1}^{j_n} \IM_{d_j^{(n)}}(\IC) \lra \bigoplus_{k=1}^{k_{n+1}}
\IM_{d_k^{(n+1)}}(\IC) ~, \\
A_1 \oplus \cdots \oplus A_{j_n} ~\mapsto~ (\oplus_{j=1}^{j_n} N_{1j} A_j) \bigoplus
\cdots \bigoplus  (\oplus_{j=1}^{j_n} N_{k_{n+1} j} A_j)~, ~~~~~~~~~~
\end{array}
\ee
where the integers $N_{kj}$ are such that  $(n, j) \searrow^{N_{kj}} (n+1, k)$ and we
have used the notation (\ref{azzo}). Notice that the dimension $d_k^{(n+1)}$ of the
factor $\IM_{d_k^{(n+1)}}(\IC)$ is not arbitrary but it is determined by a
relation like (\ref{dim}), $d_k^{(n+1)} = \sum_{j=1}^{j_n} N_{kj} d^{(n)}_j$.

\begin{figure}[t]
\begin{picture}(100,185)(6,20)
\put(30,150){\circle*{4}}
\put(30,120){\circle*{4}}
\put(30,90){\circle*{4}}
\put(30,60){\circle*{4}}
\put(60,180){\circle*{4}}
\put(60,120){\circle*{4}}
\put(60,90){\circle*{4}}
\put(60,60){\circle*{4}}
\put(90,150){\circle*{4}}
\put(90,120){\circle*{4}}
\put(90,90){\circle*{4}}
\put(90,60){\circle*{4}}
\put(30,150){\line(0,-1){30}}
\put(30,120){\line(0,-1){30}}
\put(30,90){\line(0,-1){30}}
\put(60,120){\line(0,-1){30}}
\put(60,90){\line(0,-1){30}}
\put(90,150){\line(0,-1){30}}
\put(90,120){\line(0,-1){30}}
\put(90,90){\line(0,-1){30}}
\put(30,60){\line(0,-1){10}}
\put(60,60){\line(0,-1){10}}
\put(90,60){\line(0,-1){10}}
\put(60,180){\line(1,-1){30}}
\put(30,150){\line(1,-1){30}}
\put(30,120){\line(1,-1){30}}
\put(30,90){\line(1,-1){30}}
\put(60,180){\line(-1,-1){30}}
\put(90,150){\line(-1,-1){30}}
\put(90,120){\line(-1,-1){30}}
\put(90,90){\line(-1,-1){30}}
\put(30,60){\line(1,-1){10}}
\put(90,60){\line(-1,-1){10}}
\put(54,185){{\small$1$}}
\put(16,148){{\small$1$}}
\put(16,118){{\small$1$}}
\put(16,88){{\small$1$}}
\put(16,58){{\small$1$}}
\put(97,148){{\small$1$}}
\put(97,118){{\small$1$}}
\put(97,88){{\small$1$}}
\put(97,58){{\small$1$}}
\put(63,113){$2$}
\put(63,83){$4$}
\put(63,53){$6$}
\put(45,30){$\vdots$}
\put(75,30){$\vdots$}
\end{picture}
\caption{The Bratteli diagram of the algebra $\ca_\vee$; the labels
indicate the dimension of the corresponding matrix algebras}
\label{fi:veealg}
\end{figure}
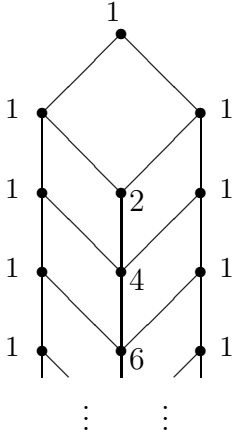

\bexam
An AF-algebra $\ca$ is commutative if and only if all the factors $\IM_{d_k^{(n)}}(\IC)$
are one dimensional, $\IM_{d_k^{(n)}}(\IC) \simeq \IC$. Thus the corresponding diagram
$\cd$ has the property that for each $(n,k) \in \cd, n > 0$, there is exactly one
$(n-1,j) \in
\cd$ such that $(n-1, j) \searrow^{1} (n, k)$.
\eexam
\bexam
Let us consider the subalgebra $\ca$ of the algebra
$\cb(\ch)$ of bounded operators on an infinite dimensional (separable) Hilbert space
$\ch = \ch_1 \oplus \ch_2$, given in the following manner. Let $\cp_j$ be the projection
operators on $\ch_j, ~j = 1, 2$, and $\ck(\ch)$ the algebra of compact operators on
$\ch$. Then, the algebra $\ca$ is
\be
\ca_\vee = \IC\cp_1 + \ck({\ch}) + \IC\cp_2~. \label{alvee}
\ee
The use of the symbol $\ca_\vee$ is due to the fact that, as we shall see below, this
algebra is associated with any part of the poset of the line in Fig.~\ref{fi:linhas},
of the form
\be
\bigvee = \{y_{i-1}, x_i, y_i\}~, \label{pvee}
\ee
in the sense that this poset is identified with the space of primitive ideals of
$\ca_\vee$. The $C^*$-algebra (\ref{alvee}) can be obtained as the inductive limit of
the following sequence of finite dimensional algebras:
\be\label{vee}
\begin{array}{l}
\ca_0 = \IM_{1}(\IC)~,   \\
\ca_1 = \IM_{1}(\IC) \oplus \IM_{1}(\IC)~,   \\
\ca_2 = \IM_{1}(\IC) \oplus \IM_{2}(\IC) \oplus \IM_{1}(\IC)~,    \\
\ca_3 = \IM_{1}(\IC) \oplus \IM_{4}(\IC) \oplus \IM_{1}(\IC)~,   \\
~~~ \vdots   \\
\ca_n = \IM_{1}(\IC) \oplus \IM_{2n-2}(\IC) \oplus \IM_{1}(\IC)~,   \\
~~~ \vdots
\end{array}
\ee
where, for $n \geq 1$, $\ca_n$ is embedded in $\ca_{n+1}$ as follows
\bea
&& \IM_{1}(\IC) \oplus \IM_{2n-2}(\IC) \oplus \IM_{1}(\IC) ~\hookrightarrow \nonumber
\\ && ~~~~~~~~~~~~~~~ \hookrightarrow ~ \IM_{1}(\IC) \oplus (\IM_{1}(\IC)
\oplus \IM_{2n-2}(\IC) \oplus \IM_{1}(\IC)) \oplus \IM_{1}(\IC)~, \nonumber \\
&& ~ \nonumber \\
&& \left[
\begin{array}{ccc}
\lambda_1 & 0  & 0    \\
0    & B  & 0    \\
0    & 0  & \lambda_2
\end{array}
\right]~~ \mapsto ~
\left[
\begin{array}{ccccc}
\lambda_1 & 0    & 0  & 0       & 0      \\
0    & \lambda_1 & 0  & 0       & 0      \\
0    & 0    & B  & 0       & 0      \\
0    & 0    & 0  & \lambda_2    & 0       \\
0    & 0    & 0  & 0       & \lambda_2
\end{array}
\right]~,
\label{vee1}
\eea
for any $\lambda_1, \lambda_2 \in \IM_{1}(\IC)$ and any $B \in \IM_{2n-2}(\IC)$.
The corresponding Bratteli diagram is shown in Fig.~\ref{fi:veealg}.
The algebra (\ref{alvee}) has three irreducible representations,
\be\label{repvee}
\begin{array}{ll}
\pi_1 : \ca_\vee \lra \cb(\ch)  ~, & a = (\lambda_1 \cp_1 + k + \lambda_2 \cp_2)
\mapsto \pi_1(a) = a~, \\
\pi_2 : \ca_\vee \lra \cb(\IC) \simeq \IC ~, & a = (\lambda_1 \cp_1 + k + \lambda_2 \cp_2)
\mapsto \pi_2(a) = \lambda_1~, \\
\pi_3 : \ca_\vee \lra \cb(\IC) \simeq \IC ~, & a = (\lambda_1 \cp_1 + k + \lambda_2 \cp_2)
\mapsto \pi_3(a) = \lambda_2~,
\end{array}
\ee
with $\lambda_1,\lambda_2 \in \IC$ and $k \in \ck(\ch)$. The corresponding kernels are
\be\label{kervee}
\ci_1 = \{0\} ~, ~~~
\ci_2 = \ck(\ch) + \IC\cp_2 ~, ~~~
\ci_3 = \IC\cp_1 + \ck(\ch) ~.
\ee
The partial order given by the inclusions $\ci_1 \subset \ci_2$ and $\ci_1 \subset
\ci_3$ (which is an equivalent way to provide the
Jacobson topology) produces a topological space $Prim{\cal A}_{\vee}$ which is just the
$\bigvee$ poset in (\ref{pvee}).
\eexam

\subsection{From Noncommutative Lattices to Bratteli Diagrams (and viceversa)}\label{se:nlb}

{}From the Bratteli diagram of an AF-algebra $\ca$ one can also obtain the
(norm closed two-sided) ideals of the latter and determine which ones are
primitive. On the set of such ideals the topology is then given by
constructing a poset whose partial order is provided by the inclusion of
ideals. Therefore, both $Prim(\ca)$ and its topology can be determined from
the Bratteli diagram of $\ca$. We refer to \cite{gianni} for details. Here
we shall briefly describe the reverse algorithm which allows one to
construct an  AF-algebra (or rather its Bratteli diagram $\cd(\ca)$) whose
primitive ideal space is a given (finitary, noncommutative) lattice $P$
\cite{Br2,BE}. We refer to \cite{ELTkappa,ELTfunctions,gianni} for more
details and several examples.
\bprop\label{most}
Let $P$ be a topological space with the following properties,
\begin{description}
\item[(i)] The space $P$ is $T_0$;
\item[(ii)] If $F \subset P$ is a closed set which is not the union of two proper
closed subsets, then $F$ is the closure of a one-point set;
\item[(iii)] The space $P$ contains at most a countable number of closed sets;
\item[(iv)] If $\{F_n\}_n$ is a decreasing ($F_{n+1} \subseteq F_n)$ sequence of closed
subsets of $P$, then $\bigcap_n F_n$ is an element in $\{F_n\}_n$.
\end{description}
Then, there exists an AF algebra $\ca$ whose primitive space $\prim$ is
homeomorphic to $P$.

\proof
The proof consists in constructing explicitly the Bratteli diagram $\cd(\ca)$ of the
algebra $\ca$. We shall sketch the main steps while referring to \cite{Br2,BE} for
more details.\\
\begin{description}
\item[$\bullet$]
Let $\{K_0, K_1, K_2, \ldots \}$ be the collection of all closed sets in the lattice
$P$, with $K_0 = P$. \\
\item[$\bullet$]
Consider the subcollection $\ck_n = \{K_0,K_1,\ldots, K_n\}$ and
let $\ck_n'$ be the smallest collection of (closed) sets in $P$ containing
$\ck_n$ which is closed under union and intersection. \\
\item[$\bullet$]
Consider the algebra of sets (We recall that a non empty
collection $R$ of subsets of a set $X$ is called an {\it algebra of sets} if $R$ is
closed under the operations of union, i.e. $E,F \in R
\Rightarrow E \cup F \in R$, and of complement, i.e.
$E \in R \Rightarrow E^c =: X \setminus E \in R$.)  generated by the collection
$\ck_n$. Then, the minimal sets $\cuai_n = \{Y_n(1), Y_n(2), \ldots, Y_n(k_n) \}$ of
this algebra form a partition of $P$. \\
\item[$\bullet$]
Let $F_n(j)$ be the smallest set in the subcollection $\ck_n'$ which contains
$Y_n(j)$. Define $\cf_n = \{F_n(1), F_n(2), \ldots, F_n(k_n)\}$. \\
\item[$\bullet$]
As a consequence of the assumptions in the Proposition one has that
\bea
&& Y_n(k) \subseteq F_n(k)~, ~~ \bigcup_k Y_n(k) = P~, ~~\bigcup_k F_n(k) = P~,
\label{a3} \\ && Y_n(k) = F_n(k) \setminus \bigcup_{p\not=k}
\{ F_n(p) ~|~ F_n(p) \subset F_n(k)\}~, \label{a4} \\
&& F_n(k) = \bigcup_{p }
\{ F_{n+1}(p) ~|~ F_{n+1}(p) \subseteq F_n(k)\}~, \label{a5}\\
&& {\rm If}~~ F\subset P ~~{\rm is ~closed}~, ~\exists ~n \geq 0~, ~{\rm s.t.}~
F_n(k) = \bigcup_{p } \{ F_{n}(p) ~|~ F_{n}(p) \subseteq F  \}. \label{a6}
\eea
\item[$\bullet$]
The diagram $\cd(\ca)$ is constructed as follows. \label{`agosto'}
\begin{enumerate}
\item[(1.)]
{\it The $n$-th level of $\cd(\ca)$ has $k_n$ points, one for each set
$Y_n(k), with k=1, \cdots, k_n$}.\\
Thus $\cd(\ca)$ is the set of all ordered pairs $(n,k), ~
k =1, \ldots, k_n, ~n = 0, 1, \ldots $~.
\item[(2.)]
{\it The point corresponding to $Y_n(k)$ at level $n$ of the diagram is linked to
the point corresponding to $Y_{n+1}(j)$ at level $n+1$, if and only if
$Y_n(k) \cap~F_{n+1}(j) \neq \emptyset$. The multiplicity of the embedding is always
$1$}. \\
Thus, the partial embeddings of the diagram are given by
\be
(n,k) ~\searrow^p~ (n+1,j)~, ~~~{\rm with}~~~
\left\{
\begin{array}{l}
p = 1 ~~{\rm if}~~ Y_n(k) \cap~F_{n+1}(j) \neq \emptyset ~, \\
p = 0 ~~{\rm otherwise}~.
\end{array}
\right.
\ee
\end{enumerate}
\end{description}
That the diagram $\cd(\ca)$ is really the diagram of an AF algebra $\ca$,
namely that conditions $(i)-(iv)$ of page~\pageref{`brdiagram'} are satisfied, follows
from the conditions (\ref{a4})-(\ref{a6}) above.
\eprop

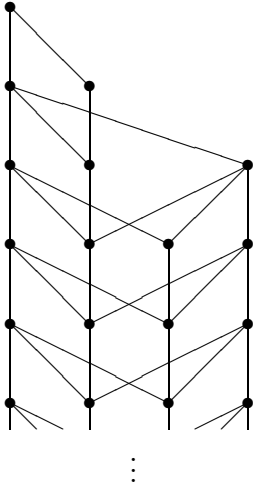
\begin{figure}[t]
\begin{picture}(110,210)(20,-10)
\put(30,180){\circle*{4}}
\put(30,150){\circle*{4}}
\put(30,120){\circle*{4}}
\put(30,90){\circle*{4}}
\put(30,60){\circle*{4}}
\put(30,30){\circle*{4}}
\put(60,150){\circle*{4}}
\put(60,120){\circle*{4}}
\put(60,90){\circle*{4}}
\put(60,60){\circle*{4}}
\put(60,30){\circle*{4}}
\put(90,90){\circle*{4}}
\put(90,60){\circle*{4}}
\put(90,30){\circle*{4}}
\put(120,30){\circle*{4}}
\put(120,60){\circle*{4}}
\put(120,90){\circle*{4}}
\put(120,120){\circle*{4}}
\put(30,180){\line(0,-1){30}}
\put(30,180){\line(1,-1){30}}
\put(30,150){\line(0,-1){30}}
\put(30,150){\line(1,-1){30}}
\put(30,150){\line(3,-1){90}}
\put(60,150){\line(0,-1){30}}
\put(30,60){\line(0,-1){30}}
\put(60,60){\line(0,-1){30}}
\put(90,60){\line(0,-1){30}}
\put(120,60){\line(0,-1){30}}
\put(30,120){\line(0,-1){30}}
\put(60,120){\line(0,-1){30}}
\put(30,90){\line(0,-1){30}}
\put(60,90){\line(0,-1){30}}
\put(90,90){\line(0,-1){30}}
\put(120,120){\line(0,-1){30}}
\put(120,90){\line(0,-1){30}}
\put(30,120){\line(1,-1){30}}
\put(30,90){\line(1,-1){30}}
\put(30,60){\line(1,-1){30}}
\put(120,120){\line(-1,-1){30}}
\put(120,90){\line(-1,-1){30}}
\put(120,60){\line(-1,-1){30}}
\put(120,120){\line(-2,-1){60}}
\put(120,90){\line(-2,-1){60}}
\put(120,60){\line(-2,-1){60}}
\put(30,120){\line(2,-1){60}}
\put(30,90){\line(2,-1){60}}
\put(30,60){\line(2,-1){60}}
\put(30,30){\line(0,-1){10}}
\put(30,30){\line(1,-1){10}}
\put(30,30){\line(2,-1){20}}
\put(60,30){\line(0,-1){10}}
\put(90,30){\line(0,-1){10}}
\put(120,30){\line(-1,-1){10}}
\put(120,30){\line(0,-1){10}}
\put(120,30){\line(-2,-1){20}}
\put(75,0){$\vdots$}
\end{picture}
\caption{The Bratteli diagram for the circle poset $P_4(S^1)$ }
\label{fi:cirbra}
\end{figure}

\noindent
We know that different algebras could yield the same space of primitive ideals
(strong Morita equivalence).
It may happen that by changing the order in which the closed sets of $P$ are taken in the
construction of the previous proposition, one produces different algebras, all
having the same space of primitive ideals though, and so all producing spaces which are
homeomorphic to the starting $P$ (any two of these spaces being, a fortiori,
homeomorphic).
\bexam
As a simple example, consider again the lattice,
\be
\bigvee = \{y_{i-1}, x_i, y_i\} \equiv \{x_2, x_1, x_3 \}~.
\ee
This topological space contains four closed sets:
\be
K_0=\{x_2, x_1, x_3 \}~, K_1 =\{x_2\}~, K_2 =\{x_3\}~, K_3 =\{x_2, x_3\} =
K_1
\cup K_2~.
\ee
Thus, with the notation of Proposition~\ref{most}, it is not difficult to check that:
\[
\begin{array}{ll}
\ck_0=\{K_0\}~,             & \ck_0'=\{K_0\}~, \\
\ck_1=\{K_0,K_1\}~,         & \ck_1'=\{K_0,K_1\}~, \\
\ck_2=\{K_0,K_1,K_2\}~,     & \ck_2'=\{K_0,K_1,K_2,K_3\}~, \\
\ck_3=\{K_0,K_1,K_2,K_3\}~, & \ck_3'=\{K_0,K_1,K_2,K_3\}~, \\
\vdots & ~
\end{array}
\]
\be
\begin{array}{lll}
Y_0(1)=\{x_1, x_2, x_3\}~,  & ~~ & F_0(1)=K_0~,          \\
~ & ~ & ~  \\
Y_1(1)=\{x_2\}~,~~~~~ Y_1(2)=\{x_1, x_3\}~,
                                     & ~~ & F_1(1)=K_1~,~~~~~ F_1(2)=K_0~, \\
~ & ~ & ~  \\
Y_2(1)=\{x_2\}~,~~~~~ Y_2(2)=\{x_1\}~,
                                     & ~~ & F_2(1)=K_1~,~~~~~ F_2(2)=K_0~, \\
Y_2(3)=\{x_3\}~,                     & ~~ & F_2(3)=K_2~,                   \\
~ & ~ & ~  \\
Y_3(1)=\{x_2\}~,~~~~~ Y_3(2)=\{x_1\}~,
                                     & ~~ & F_3(1)=K_1~,~~~~~ F_3(2)=K_0~, \\
Y_3(3)=\{x_3\}~,                     & ~~ & F_3(2)=K_2~,                   \\
\vdots                               & ~~ & ~
\end{array}
\ee
Since $\bigvee$ has only a finite number of points (three), and hence a finite number
of closed sets (four), the partition of $\bigvee$ repeats itself after the third
level.
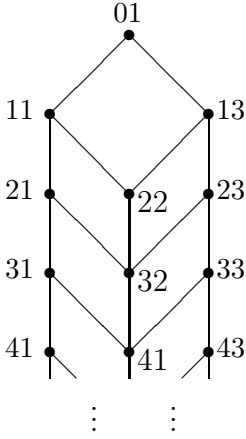
\begin{figure}[t]
\begin{picture}(100,185)(3,20)
\put(30,150){\circle*{4}}
\put(30,120){\circle*{4}}
\put(30,90){\circle*{4}}
\put(30,60){\circle*{4}}
\put(60,180){\circle*{4}}
\put(60,120){\circle*{4}}
\put(60,90){\circle*{4}}
\put(60,60){\circle*{4}}
\put(90,150){\circle*{4}}
\put(90,120){\circle*{4}}
\put(90,90){\circle*{4}}
\put(90,60){\circle*{4}}
\put(30,150){\line(0,-1){30}}
\put(30,120){\line(0,-1){30}}
\put(30,90){\line(0,-1){30}}
\put(60,120){\line(0,-1){30}}
\put(60,90){\line(0,-1){30}}
\put(90,150){\line(0,-1){30}}
\put(90,120){\line(0,-1){30}}
\put(90,90){\line(0,-1){30}}
\put(30,60){\line(0,-1){10}}
\put(60,60){\line(0,-1){10}}
\put(90,60){\line(0,-1){10}}
\put(60,180){\line(1,-1){30}}
\put(30,150){\line(1,-1){30}}
\put(30,120){\line(1,-1){30}}
\put(30,90){\line(1,-1){30}}
\put(60,180){\line(-1,-1){30}}
\put(90,150){\line(-1,-1){30}}
\put(90,120){\line(-1,-1){30}}
\put(90,90){\line(-1,-1){30}}
\put(30,60){\line(1,-1){10}}
\put(90,60){\line(-1,-1){10}}
\put(54,185){{\small$01$}}
\put(13,148){{\small$11$}}
\put(13,118){{\small$21$}}
\put(13,88){{\small$31$}}
\put(13,58){{\small$41$}}
\put(93,148){{\small$13$}}
\put(93,118){{\small$23$}}
\put(93,88){{\small$33$}}
\put(93,58){{\small$43$}}
\put(63,113){$22$}
\put(63,83){$32$}
\put(63,53){$41$}
\put(45,30){$\vdots$}
\put(75,30){$\vdots$}
\end{picture}
\caption{The Bratteli diagram associated with the poset $\bigvee$; the
label $nk$ stands for $Y_n(k)$ }
\label{fi:bigbra}
\end{figure}
Figure~\ref{fi:bigbra} shows the corresponding diagram, obtained through rules (1.) and
(2.) in Proposition~\ref{most} above (on page~\pageref{`agosto'}). By using the
fact that the first matrix algebra
$\ca_0$ is $\IC$ and the fact that all the embeddings have multiplicity one, the diagram
of Fig.~\ref{fi:bigbra} is seen to coincide with the diagram of Fig.~\ref{fi:veealg}.
As we have previously said, the latter corresponds to the AF-algebra
\be
\ca_\vee =
\IC\cp_1 + \ck({\ch}) + \IC\cp_2~, ~ \ch = \ch_1 \oplus \ch_2~.
\ee
\eexam
\bexam
Another interesting example is provided by the lattice $P_4(S^1)$ for the
one-dimensional sphere in Fig.~\ref{fi:cirhas}.
This topological space contains six closed sets:
\be
\begin{array}{l}
K_0 = \{x_1, x_2, x_3, x_4 \}~, ~K_1 = \{x_1, x_3, x_4 \}~, ~K_2 =\{x_3\}~,
~K_3 =\{x_4\}~,   \\
K_4 = \{x_2, x_3, x_4 \}~, ~K_5 = \{x_3, x_4 \} = K_2 \cup K_3~.
\end{array}
\ee
Thus, with the notation of Proposition~\ref{most}, one finds,
\be
\begin{array}{lll}
\ck_0=\{K_0\}~,                      & ~ &  \ck_0'=\{K_0\}~, \\
\ck_1=\{K_0,K_1\}~,                  & ~ &  \ck_1'=\{K_0,K_1\}~, \\
\ck_2=\{K_0,K_1,K_2\}~,              & ~ &  \ck_2'=\{K_0,K_1,K_2\}~,  \\
\ck_3=\{K_0,K_1,K_2,K_3\}~,          & ~ &  \ck_3'=\{K_0,K_1,K_2,K_3,K_5\}~,  \\
\ck_4=\{K_0,K_1,K_2,K_3,K_4\}~,      & ~ &  \ck_4'=\{K_0,K_1,K_2,K_3,K_4,K_5\}~, \\
\ck_5=\{K_0,K_1,K_2,K_3,K_4,K_5\}~,  & ~ &  \ck_5'=\{K_0,K_1,K_2,K_3,K_4,K_5\}~, \\
\vdots & & \\
~ & ~ & ~ \\
Y_0(1)=\{x_1, x_2, x_3, x_4\}~,  & ~~ & F_0(1)=K_0~,   \\
~ & ~ & ~  \\
Y_1(1)=\{x_1, x_3, x_4\}~,
                                 & ~~ & ~              \\
~~~~~~~~~~~~~~~~~~~~~~~~ Y_1(2)=\{x_2\}~,
                                     & ~~ & F_1(1)=K_1~,~~~~~ F_1(2)=K_0~,  \\
~ & ~ & ~   \\
Y_2(1)=\{x_3\}~,~~~~~ Y_2(2)=\{x_2\}~,
                                     & ~~ & F_2(1)=K_2~,~~~~~ F_2(2)=K_0~,  \\
Y_2(3)=\{x_1, x_4\}~,  ~
                                     & ~~ & F_2(3)=K_1~,  ~            \\
~ & ~ & ~   \\
 Y_3(1)=\{x_3\}~,~~~~~ Y_3(2)=\{x_2\}~,
                                     & ~~ & F_3(1)=K_2~,~~~~~ F_3(2)=K_0~, \\
Y_3(3)=\{x_1\}~,~~~~~ Y_3(4)=\{x_4\}~,
                                     & ~~  & F_3(3)=K_1~,~~~~~ F_3(4)=K_3~, \\
~ & ~ & ~   \\
Y_4(1)=\{x_3\}~,~~~~~ Y_4(2)=\{x_2\}~,
                                     & ~~ & F_4(1)=K_2~,~~~~~ F_4(2)=K_4~, \\
Y_4(3)=\{x_1\}~,~~~~~ Y_4(4)=\{x_4\}~,
                                     & ~~ & F_4(3)=K_1~,~~~~~ F_4(4)=K_3~, \\
~ & ~ & ~   \\
Y_5(1)=\{x_3\}~,~~~~~ Y_5(2)=\{x_2\}~,
                                     & ~~ & F_5(1)=K_2~,~~~~~ F_5(2)=K_4~, \\
Y_5(3)=\{x_1\}~,~~~~~ Y_5(4)=\{x_4\}~,
                                     & ~~ & F_5(3)=K_1~,~~~~~ F_5(4)=K_3~, \\
\vdots &  &
\end{array}
\ee
Since there are a finite number of points (four), and hence a finite number
of closed sets (six), the partition of $P_4(S^1)$ repeats itself after the fourth
level. The corresponding Bratteli diagram is exhibited in Fig.~\ref{fi:cirbra}.
The ideal $\{0\}$ is not primitive.
The algebra is given by
\be\label{cirafl}
\begin{array}{l}
\ca_0 = \IM_{1}(\IC)~,   \\
\ca_1 = \IM_{1}(\IC) \oplus \IM_{1}(\IC)~,    \\
\ca_2 = \IM_{1}(\IC) \oplus \IM_{2}(\IC) \oplus \IM_{1}(\IC)~, \\
\ca_3 = \IM_{1}(\IC) \oplus \IM_{4}(\IC) \oplus \IM_{2}(\IC) \oplus
\IM_{1}(\IC)~,  \\
\ca_4 = \IM_{1}(\IC) \oplus \IM_{6}(\IC) \oplus \IM_{4}(\IC) \oplus
\IM_{1}(\IC)~,  \\
 ~~~ \vdots \\
\ca_n = \IM_{1}(\IC) \oplus \IM_{2n-2}(\IC) \oplus \IM_{2n-4}(\IC) \oplus
\IM_{1}(\IC)~,  \\
~~~ \vdots
\end{array}
\ee
where, for $n > 2$, $\ca_n$ is embedded in $\ca_{n+1}$ as follows
\be\left[
\begin{array}{cccc}
\lambda_1 &   &   &   \\
     & B &   &  \\
     &   & C &   \\
     &   &   & \lambda_2
\end{array}
\right]~~ \mapsto ~
\left[
\begin{array}{cccccccc}
\lambda_1 &      &     &      &       &    &      &   \\
     & \lambda_1 & 0   & 0    &       &    &      &   \\
     & 0    & B   & 0    &       &    &      &   \\
     & 0    & 0   & \lambda_2 &       &    &      &   \\
     &      &     &      & \lambda_1  & 0  & 0    &   \\
     &      &     &      & 0     & C  & 0    &   \\
     &      &     &      & 0     & 0  & \lambda_2 &   \\
     &      &     &      &       &    &      & \lambda_2
\end{array}
\right]~,
\label{cirafl1}
\ee
with $\lambda_1, \lambda_2 \in \IM_{1}(\IC)$, $B \in \IM_{2n-2}(\IC)$ and $C \in
\IM_{2n-4}(\IC)$; elements which are not shown are equal to zero. The algebra
limit $\ca_{P_4(S^1)}$ can be realized explicitly as  a subalgebra of bounded
operators on an infinite dimensional Hilbert space $\ch$ naturally associated
with the poset $P_4(S^1)$. Firstly, to any {\it link} $(x_i, x_j), x_i \succ
x_j,$  of the poset one associates a Hilbert space
$\ch_{ij}$; for the case at hand, one has four Hilbert spaces, $\ch_{31}, \ch_{32},
\ch_{41}, \ch_{42}$. Then, since all links are at the same level, $\ch$ is just given by
the direct sum
\be
\ch = \ch_{31} \oplus \ch_{32} \oplus \ch_{41} \oplus \ch_{42}~.
\ee
The algebra $\ca_{P_4(S^1)}$ is given by \cite{ELTfunctions},
\be
\ca_{P_4(S^1)} = \IC\cp_{\ch_{31} \oplus \ch_{32}} + \ck_{\ch_{31} \oplus \ch_{41}}
+ \ck_{\ch_{32} \oplus \ch_{42}} + \IC\cp_{\ch_{41} \oplus \ch_{42}}~.
\label{ciralg}
\ee
Here $\ck$ denotes compact operators and $\cp$ orthogonal projection.
The algebra (\ref{ciralg}) has four irreducible representations. Any element
$a\in\ca_{P_4(S^1)}$ is of the form
\be
a = \lambda \cp_{3,12} + k_{34,1} + k_{34,2} + \mu\cp_{4,12}~,
\ee
with $\lambda,\mu \in \IC$, $k_{34,1}\in\ck_{\ch_{31} \oplus \ch_{41}}$ and
$k_{34,2}
\in \ck_{\ch_{32} \oplus \ch_{42}}$. The representations are,
\be\label{repcir}
\begin{array}{ll}
\pi_1 : \ca_{P_4(S^1)} \lra \cb(\ch) ~,
& a \mapsto \pi_1(a) =  \lambda \cp_{3,12} + k_{34,1} + \mu\cp_{4,12}~,  \\
\pi_2 : \ca_{P_4(S^1)} \lra \cb(\ch)  ~,
& a \mapsto \pi_2(a) =
\lambda \cp_{3,12} + k_{34,2} + \mu\cp_{4,12}~,    \\
\pi_3 : \ca_{P_4(S^1)} \lra \cb(\IC) \simeq \IC ~, & a \mapsto \pi_3(a) =
\lambda~,
\nonumber \\
\pi_4 : \ca_{P_4(S^1)} \lra \cb(\IC) \simeq \IC ~, & a \mapsto \pi_4(a) =
\mu~,
\end{array}
\ee
with corresponding kernels,
\bea\label{kercir}
&&\ci_1 = \ck_{\ch_{32} \oplus \ch_{42}}~, ~~
\ci_2 = \ck_{\ch_{31} \oplus \ch_{41}}~, ~~ \nonumber \\
&&\ci_3 = \ck_{\ch_{31} \oplus \ch_{41}} + \ck_{\ch_{32} \oplus \ch_{42}} +
                         \IC\cp_{\ch_{41} \oplus \ch_{42}}~, \nonumber \\
&&\ci_4 = \IC\cp_{\ch_{31} \oplus \ch_{32}} + \ck_{\ch_{31} \oplus \ch_{41}}
                          + \ck_{\ch_{32} \oplus \ch_{42}}~.
\eea
The partial order given by the inclusions $\ci_1 \subset \ci_3$, $\ci_1 \subset \ci_4$
and $\ci_2 \subset \ci_3$, $\ci_2 \subset \ci_4$ produces a topological space
$Prim{\cal A}_{P_4(S^1)}$ which is just the circle poset in Fig.~\ref{fi:cirhas}.
\eexam

\subsection{The general case}
In fact, by looking at the previous examples a bit more
carefully one can infer the algorithm by which one goes from a (finite) poset $P$ to the
corresponding Bratteli diagram $\cd(\ca_P)$. Let $(x_1,
\cdots , x_N)$ be the points of
$P$ and for
$k=1,
\cdots, N$, let $S_k =: \bar{\{x_k \}}$ be the smallest closed subset of $P$ containing
the point $ x_j$. Then, the Bratteli diagram repeats itself after level $N$ and the
partition $Y_n(k)$ of Proposition~\ref{most} is just given by
\be
Y_n(k) = Y_{n+1}(k) = \{x_k \}~, ~~k = 1, \dots, N~, ~~~\forall ~n \geq N~.
\ee
As for the associated $F_n(k)$ , from level $N+1$ on, they are
given by the
$S_k$,
\be
F_n(k) = F_{n+1}(k) = S_k~, ~k = 1, \dots, N~, ~~~\forall ~n \geq N+1~.
\ee
In the diagram $\cd(\ca_P)$, for any $n \geq N$, $(n,k) \searrow (n+1,j)$ if and only
if $\{x_k \} \bigcap S_j \not= \emptyset$, that is if and only if $x_k \in S_j$. \\
We also sketch the algorithm used to construct the algebra limit
$\ca_P$
determined by the Bratteli diagram $\cd(\ca_P)$ (This algebra is really defined
only modulo Morita equivalence).
\cite{BL,ELTfunctions}. The idea is to associate to the poset $P$ an infinite
dimensional separable Hilbert space
$\ch(P)$ out of tensor products and direct sums of infinite dimensional
(separable) Hilbert spaces $\ch_{ij}$ associated with each link $(x_i, x_j), x_i \succ
x_j$, in the poset. (The Hilbert spaces could all be taken to be the same. The
label is there just to distinguish among them.)
Then for each point
$x \in P$ there is a subspace $\ch(x) \subset \ch(P)$ and an algebra $\cb(x)$ of bounded
operators acting on $\ch(x)$. The algebra $\ca_P$ is the one generated by all the
$\cb(x)$ as
$x$ varies in $P$. In fact, the algebra $\cb(x)$ can be made to act on the whole of
$\ch(P)$ by defining its action on the complement of $\ch(x)$ to be zero.
Consider any {\it maximal chain} $C_\alpha$ in $P$:
$C_\alpha =
\{x_\alpha, \dots, x_2, x_1 ~|~ x_j
\succ x_{j-1}\}$ for any maximal point $x_\alpha \in P$. To this chain one
associates the Hilbert  space
\be
\ch(C_\alpha) = \ch_{\alpha,\alpha-1} \otimes \cdots \otimes \ch_{3,2} \otimes
\ch_{2,1}~.
\ee
By taking the direct sum over all maximal chains, one gets the Hilbert space $\ch(P)$,
\be\label{htot}
\ch(P) = \bigoplus_\alpha \ch(C_\alpha)~.
\ee
The subspace $\ch(x) \subset \ch(P)$ associated with any point $x \in P$ is
constructed in a similar way by restricting the sum to all maximal chains containing
the point $x$. It can be split into two parts,
\be\label{fibhil}
\ch(x) = \ch(x)^u \otimes \ch(x)^d~,
\ee
with,
\be
\begin{array}{l}
\ch(x)^u = \ch(P^u_x)~, ~~~ P^u_x = \{y \in P ~|~ y \succeq x\}~, \\
\ch(x)^d = \ch(P^d_x)~, ~~~ P^d_x = \{y \in P ~|~ y \preceq x\}~.
\end{array}
\ee
Here $\ch(P^u_x)$ and $\ch(P^d_x)$ are constructed as in (\ref{htot}); also,
$\ch(x)^u =\IC$ if $x$ is a maximal point and $\ch(x)^d =\IC$ if $x$ is a minimal point.
Consider now the algebra
$\cb(x)$ of bounded operators on
$\ch(x)$ given by
\be\label{fibalg}
\cb(x) = \ck(\ch(x)^u) \otimes \IC \cp(\ch(x)^d) \simeq \ck(\ch(x)^u) \otimes
\cp(\ch(x)^d)~.
\ee
As before, $\ck$ denotes compact operators and $\cp$ orthogonal projection. We see
that  $\cb(x)$ acts by compact operators on the Hilbert space $\ch(x)^u$ determined by
the points which follow $x$ and by multiples of the identity on the Hilbert space
$\ch(x)^d$ determined by the points which precede $x$. These algebras satisfy the
rules: $\cb(x) \cb(y) \subset \cb(x)$ if $x\preceq y$ and $\cb(x) \cb(y) = 0$ if $x$ and
$y$ are not comparable. As already mentioned, the algebra $\ca(P)$ of the poset $P$ is the algebra
of bounded operators on $\ch(P)$ generated by all $\cb(x)$ as $x$ varies over $P$. It
can be shown that $\ca(P)$ has a space of primitive ideals which is homeomorphic to
the poset $P$ \cite{BL,ELTfunctions}. We refer to \cite{ELTkappa,ELTfunctions} for
additional details and examples.

\subsection{Recovering the Algebra}\label{se:raba}

In Sect.~\ref{se:recospa} we have described how to recover a topological space $M$ in
the limit, by considering a sequence of finer and finer coverings of $M$. We constructed
a projective system of finitary topological spaces and continuous maps $\{P_i, \pi_{ij}
\}_{i,j \in \IN}$ associated with the coverings; the maps $\pi_{ij} : P_j \ra P_i~, ~j
\geq i$, being continuous surjections. The limit of the system is a topological space
$P_\infty$, in which  $M$ is embedded as the subspace of closed points. On
each point $m$ of (the image of) $M$ there is a fibre of `extra points';
the latter are all points of $P_\infty$ which `cannot be separated' by $m$.

>From a dual point of view we get a
{\it inductive system of algebras and homomorphisms}
$\{\ca_i, \phi_{ij} \}_{i,j \in \IN}$; the maps $\phi_{ij} : \ca_i \ra \ca_j~, ~j
\geq i$, being injective homeomorphisms. The system has a unique
{\it inductive limit}
$\ca^\infty$. Each algebra $\ca_i$ is such that $\ha_i = P_i$ and is associated with
$P_i$ as  described previously, $\ca_i = \ca(P_i)$. The map $\phi_{ij}$ is a `suitable
pullback' of the corresponding surjection $\pi_{ij}$. The limit space $P_\infty$ is the
structure space of the limit algebra $\ca^\infty$, $P_\infty = \ha^\infty$. And, finally
the algebra $C(M)$ of continuous functions on $M$ can be identified with the {\it center}
of
$\ca^\infty$.

We also get a inductive system of Hilbert spaces together with isometries
$\{\ch_i, \tau_{ij} \}_{i,j \in \IN}$; the maps
$\tau_{ij} : \ch_i \ra \ch_j~, ~j \geq i$, being injective isometries onto the image.
The system has a unique  {\it inductive limit}
$\ch^\infty$. Each Hilbert space $\ch_i$ is associated with the space $P_i$ as in
(\ref{htot}), $\ch_i = \ch(P_i)$, the algebra $\ca_i$ being the corresponding subalgebra
of bounded operators. The maps $\tau_{ij}$ are constructed out of the corresponding
$\phi_{ij}$. The limit Hilbert space $\ch^\infty$ is associated with the space
$P_\infty$ as in (\ref{htot}), $\ch^\infty = \ch(P_\infty)$, the algebra $\ca^\infty$
again being the corresponding subalgebra of bounded operators. And, finally, the
Hilbert space $L^2(M)$ of square integrable functions is `contained' in
$\ch^\infty$ : $\ch^\infty = L^2(M) \oplus_\alpha \ch_\alpha$, the sum being on
the `extra points' in $P_\infty$.

All of the previous is described in great details in \cite{pangs}. Here we only make a
few additional remarks. By improving the approximation (by increasing the number of
`detectors') one gets a noncommutative lattice whose Hasse diagram has a
bigger  number of points and links. The associated Hilbert space gets `more refined' :
one may think of a {\it unique (and the same)} Hilbert space which is being refined while
being  split by means of tensor products and direct sums. In the limit the information is
enough to recover completely the Hilbert space (in fact, to recover more than it).
Further considerations along these lines and possible applications to quantum mechanics
will have to await another occasion.

\subsection{Operator Valued Functions on Noncommutative Lattices}\label{se:ovf}

Much in the
same way as it happens for the commutative algebras \cite{FD},
elements of a noncommutative $C^*$-algebra whose primitive spectrum $\prim$ is a
noncommutative lattice can be realized as operator-valued functions on
$\prim$. The value of $a\in\ca$ at the
`point' $\ci\in \prim$ is just the image of $a$ under the representation $\pi_\ci$
associated with $\ci$ and such that $\ker(\pi_\ci) = \ci$,
\be
a(\ci) = \pi_\ci(a) \simeq a/\ci~, ~~~ \forall ~a \in \ca, ~\ci \in \prim~.
\ee
\begin{figure}
\begin{picture}(100,113)(-55,-43)
\put(-30,30){\circle*{4}}
\put(30,30){\circle*{4}}
\put(0,-30){\circle*{4}}
\put(-30,30){\line(1,-2){30}}
\put(30,30){\line(-1,-2){30}}
\put(-45,29){$\lambda_1$}
\put(35,29){$\lambda_2$}
\put(5,-33){$\lambda_1\cp_1 + k_{12} + \lambda_2\cp_2$}
\put(-45,50){$a = \lambda_1\cp_1 + k_{12} + \lambda_2\cp_2$}
\end{picture}
\caption{A function over the lattice $\bigvee$ }
\label{fi:veefun}
\end{figure}
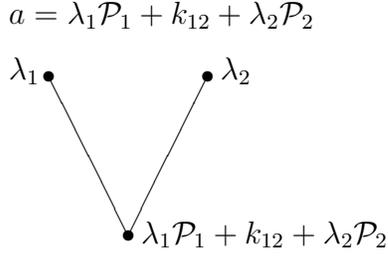
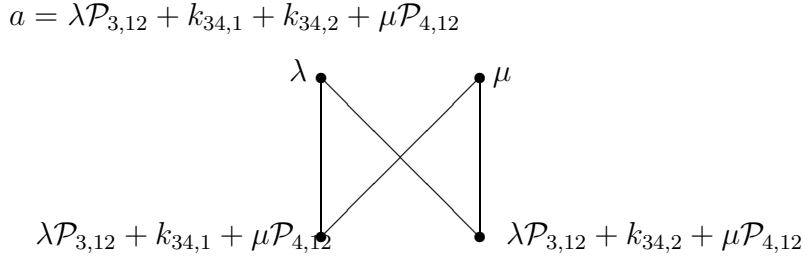
\begin{figure}
\begin{picture}(220,113)(-158,-43)
\put(-30,30){\circle*{4}}
\put(30,30){\circle*{4}}
\put(-30,-30){\circle*{4}}
\put(30,-30){\circle*{4}}
\put(-30,30){\line(0,-1){60}}
\put(30,30){\line(0,-1){60}}
\put(-30,30){\line(1,-1){60}}
\put(30,30){\line(-1,-1){60}}
\put(-42,29){$\lambda$}
\put(35,29){$\mu$}
\put(-138,-33){$\lambda\cp_{3,12} + k_{34,1} + \mu\cp_{4,12}$}
\put(40,-33){$\lambda\cp_{3,12} + k_{34,2} + \mu\cp_{4,12}$}
\put(-148,50){$a = \lambda \cp_{3,12} + k_{34,1} + k_{34,2} + \mu\cp_{4,12}$}
\end{picture}
\caption{A function over the lattice $P_4(S^1)$ }
\label{fi:cirfun}
\end{figure}
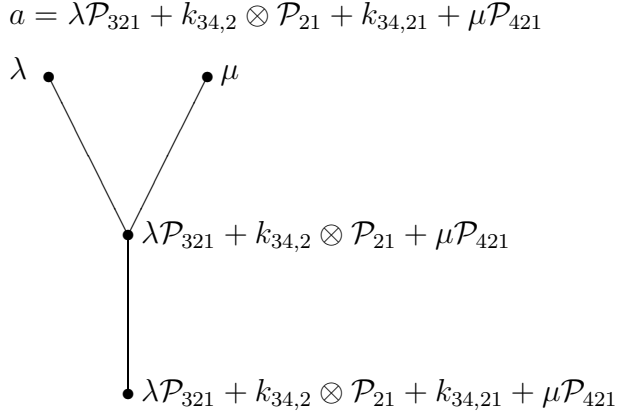
\begin{figure}
\begin{picture}(100,173)(-55,-103)
\put(-30,30){\circle*{4}}
\put(30,30){\circle*{4}}
\put(0,-30){\circle*{4}}
\put(0,-90){\circle*{4}}
\put(-30,30){\line(1,-2){30}}
\put(30,30){\line(-1,-2){30}}
\put(0,-30){\line(0,-1){60}}
\put(-45,29){$\lambda$}
\put(35,29){$\mu$}
\put(5,-33){$\lambda \cp_{321} + k_{34,2} \otimes \cp_{21} + \mu \cp_{421}$}
\put(5,-93){$\lambda \cp_{321} + k_{34,2} \otimes \cp_{21} + k_{34,21} + \mu
\cp_{421}$}
\put(-45,50){$a = \lambda \cp_{321} + k_{34,2} \otimes \cp_{21} + k_{34,21} +
\mu
\cp_{421}$}
\end{picture}
\caption{A function over the lattice $Y$ }
\label{fi:ipsfun}
\end{figure}
All this is shown pictorially in Figs.~ \ref{fi:veefun}, \ref{fi:cirfun} and
\ref{fi:ipsfun} for the $\bigvee$ lattice, a circle lattice and a lattice $Y$,
respectively.  As it is evident in those Figures, the values of a function at points
which cannot be separated by the topology differ by a compact operator. This is an
illustration of the fact that compact operators play the r\^ole of `infinitesimals'
as is discussed at length in \cite{Co1}. Furthermore, while in Figs.~\ref{fi:veefun}
and \ref{fi:cirfun} we have only `infinitesimals of first order', for the three level
lattice of Fig.~\ref{fi:ipsfun} we have both infinitesimals of first order, like
$k_{34,2}$, and infinitesimals of second order, like $k_{34,21}$.

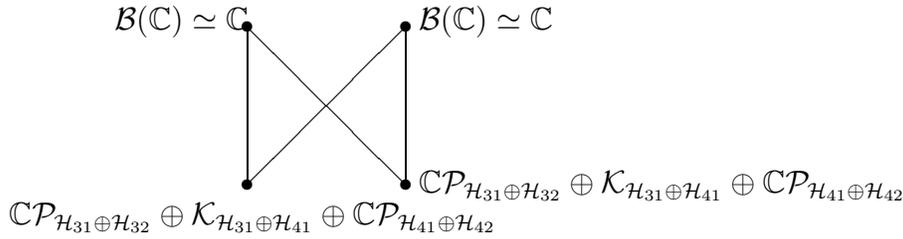
\begin{figure}
\begin{picture}(200,105)(-130,-55)
\put(-30,30){\circle*{4}}
\put(30,30){\circle*{4}}
\put(-30,-30){\circle*{4}}
\put(30,-30){\circle*{4}}
\put(-30,30){\line(0,-1){60}}
\put(30,30){\line(0,-1){60}}
\put(-30,30){\line(1,-1){60}}
\put(30,30){\line(-1,-1){60}}
\put(-80,29){$\cb(\IC)\simeq\IC$}
\put(35,29){$\cb(\IC)\simeq\IC$}
\put(-120,-45){$\IC\cp_{\ch_{31} \oplus \ch_{32}} \oplus \ck_{\ch_{31} \oplus
\ch_{41}} \oplus \IC\cp_{\ch_{41} \oplus \ch_{42}}$}
\put(35,-33){$\IC\cp_{\ch_{31} \oplus \ch_{32}} \oplus \ck_{\ch_{31} \oplus \ch_{41}}
\oplus \IC\cp_{\ch_{41} \oplus \ch_{42}}$}
\end{picture}
\caption{The fibres of the trivial bundle over the lattice $P_4(S^1)$ }
\label{fi:cirbun}
\end{figure}

In fact \cite{Co1}, the correct way of thinking of any noncommutative
$C^*$-algebra $\ca$ is as the module of sections of the `rank one trivial
vector bundle' over the associated noncommutative space. For the kind of
noncommutative lattices we are interested in, it is possible to explicitly
construct the bundle over the lattice. Such bundles are examples of {\it
bundles of $C^*$-algebras}, the fibre over any point $\ci \in \prim$ being
just the algebra of bounded operators $\pi_\ci(\ca) \subset
\cb(\ch_\ci)$, with $\ch_\ci$ the representation space. The Hilbert space and the algebra
are given explicitly by the Hilbert space in (\ref{fibhil}) and the algebra in
(\ref{fibalg})   respectively, by taking for $x$ the point
$\ci$. (At the same time, one is also constructing a bundle of Hilbert spaces.)
It is also possible to endow
the total space with a topology in such a manner that elements of
$\ca$ are realized as continuous sections.  Figure~\ref{fi:cirbun} shows the trivial bundle
over the lattice
$P_4(S^1)$.

\section{$\theta$-Angles on Noncommutative Lattices}\label{se:qmm}
As a very simple example of a
quantum mechanical system which can be studied with the techniques of
noncommutative geometry on noncommutative lattices, we shall construct the
$\theta$-quantization of
a particle on a lattice for the circle. We shall do so by constructing an
appropriate `line bundle' with a connection. We refer to \cite{ncl} and
\cite{bbllt} for more details and additional field theoretical examples. In
particular, in
\cite{bbllt} Wilson's actions for gauge and fermionic fields and
analogues of topological and Chern-Simons actions were derived.

The real line $\IR^1$ is the universal covering space of the circle $S^1$,
and the fundamental group $\pi_1(S^1) = \IZ$ acts on $\IR^1$ by translation
$
\IR^1 \ni x \rightarrow x + N ~ , ~ N \in \IZ .
$
The quotient space of this action is $S^1$ and the projection
$ : \IR^1 \ra S^1$ is given by $ \IR^1 \ni x \ra e^{i2\pi x} \in S^1$.
The domain of a typical Hamiltonian $H$ for a particle on $S^1$ need
not consist of functions on $S^1$. Rather it can be obtained from
functions $\psi_{\theta}$ on $\IR^1$ transforming under an irreducible
representation of $\pi(S^1) = \IZ$,
$
\rho_{\theta} : N \rightarrow e^{iN\theta}
$
according to
$
\psi_{\theta}(x+N) = e^{iN\theta} \psi_{\theta}(x).
$
The domain $D_{\theta}(H)$ for a typical Hamiltonian $H$ then consists of these
$\psi_{\theta}$ restricted to a fundamental domain $0 \leq x \leq 1$ for the
action of $\IZ$, and subject to a differentiability requirement:
\be
D_{\theta}(H) = \{ \psi_{\theta} : \psi_{\theta}(1) = e^{i\theta}
\psi_{\theta}(0)~;~~
\frac{d\psi_{\theta}(1)}{dx} = e^{i\theta} \frac{d\psi_{\theta}(0)}{dx} \} ~.
\label{dom}
\ee
In addition, $H\psi_{\theta}$ must be square integrable.
One obtains a distinct quantization, called $\theta$-quantization, for each
choice of $e^{i\theta}$.

Equivalently, wave functions can be taken to be  single-valued functions on
$S^1$ while adding a `gauge potential' term to the Hamiltonian. To be more
precise, one constructs a line bundle over $S^1$ with a connection one-form
given by $i\theta d x$. If the Hamiltonian with the domain (\ref{dom}) is
$-d^2 / d x^2$, then the Hamiltonian with the domain $D_0(h)$
consisting of single valued wave functions is $-(d / d x + i \theta)^2$.

There are similar quantization possibilities for a noncommutative lattice for
the circle as well \cite{ncl}. One constructs the algebraic analogue of the
trivial bundle on the lattice endowed with a gauge connection which is such
that the corresponding Laplacian has an approximate spectrum reproducing the
`continuum' one in the limit.

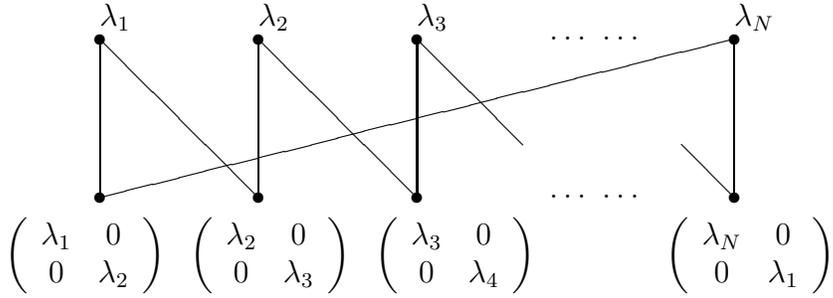
\begin{figure}[t]
\begin{picture}(340,130)(-80,-70)
\put(-30,30){\circle*{4}}
\put(30,30){\circle*{4}}
\put(90,30){\circle*{4}}
\put(210,30){\circle*{4}}
\put(-30,-30){\circle*{4}}
\put(30,-30){\circle*{4}}
\put(90,-30){\circle*{4}}
\put(210,-30){\circle*{4}}
\put(140,-30){$\dots$}
\put(160,-30){$\dots$}
\put(140,30){$\dots$}
\put(160,30){$\dots$}
\put(-30,-30){\line(4,1){240}}
\put(-30,30){\line(0,-1){60}}
\put(30,30){\line(0,-1){60}}
\put(90,30){\line(0,-1){60}}
\put(210,30){\line(0,-1){60}}
\put(-30,30){\line(1,-1){60}}
\put(30,30){\line(1,-1){60}}
\put(90,30){\line(1,-1){40}}
\put(210,-30){\line(-1,1){20}}
\put(-30,35){$\lambda_1$}
\put(30,35){$\lambda_2$}
\put(90,35){$\lambda_3$}
\put(210,35){$\lambda_N$}
\put(-70,-55){
$\left(\begin{array}{cc}
 \lambda_1 & 0 \\
0 & \lambda_2
\end{array}
\right)$}
\put(0,-55){
$\left(\begin{array}{cc}
 \lambda_2 & 0 \\
0 & \lambda_3
\end{array}
\right)$}
\put(70,-55){
$\left(\begin{array}{cc}
 \lambda_3 & 0 \\
0 & \lambda_4
\end{array}
\right)$}
\put(180,-55){
$\left(\begin{array}{cc}
 \lambda_N & 0 \\
0 & \lambda_1
\end{array}
\right)$}
\end{picture}
\caption{$P_{2N}(S^1)$ for the approximate algebra $\cc(\ca)$ }
\label{fi:appcir2n}
\end{figure}

As we have seen in Sect.~\ref{se:toap}, the algebra $\ca$ associated with any
noncommutative lattice of the circle is rather complicated and involves
infinite dimensional operators on direct sums of infinite dimensional
Hilbert spaces. In turn, this algebra $\ca$, as it is AF (approximately finite
dimensional), can indeed be approximated by algebras of matrices. The simplest
approximation is just a commutative algebra $\cc(\ca)$ of the form
\be
\cc(\ca) \simeq \IC^N = \{ c=(\lambda_1 ,\lambda_2, \cdots, \lambda_N ) ~, \lambda_i \in \IC \}~.
\label{calg}
\ee
The algebra (\ref{calg}) can produce a noncommutative lattice
with $2N$ points by considering a particular class of not necessarily
irreducible representations as in Fig.~\ref{fi:appcir2n}.
In that Figure, the top points correspond to the irreducible one
dimensional representations
\be
\pi_i : \cc(\ca) \ra \IC~, ~~~ c \mapsto \pi_i(c) = \lambda_i~, ~~i = 1, \cdots, N~.
\ee
The bottom points correspond to the reducible two dimensional
representations
\be
\pi_{i+N} : \cc(\ca) \ra \IM_{2}(\IC)~, ~~~ c \mapsto \pi_{i+N}(c) =
\left(
\begin{array}{cc}
\lambda_i & 0 \\
0 & \lambda_{i+1}
\end{array}
\right)~, ~~i = 1, \cdots, N~,
\ee
with the additional condition that $\pi_{N+1} = \pi_1$ and
$\lambda_{n+1}=\lambda_1$. The partial order, or
equivalently the topology, is determined by the inclusion of the corresponding
kernels as in Sect.~\ref{se:toap}.

By comparing Fig.~\ref{fi:appcir2n} with the corresponding
Fig.~\ref{fi:cirfun}, we see that by trading $\ca$ with $\cc(\ca)$, all
compact operators have been put to zero. A better approximation is
obtained by approximating compact operators with finite dimensional matrices
of increasing rank.

The finite projective module of sections $\ce$ associated with the
`trivial line bundle' is just
$\cc(\ca)$ itself:
\be
\ce = \IC^N = \{ \eta = (\mu_1 ,\mu_2, \cdots, \mu_N ) ~, \mu_i \in \IC \}~.
\label{cmod}
\ee
The action of $\cc(\ca)$ on $\ce$ is simply given by
\be
\ce \times \cc(\ca) \ra \ce~, ~~~(\eta, c)
\mapsto \eta c = (\eta_1 \lambda_1, \eta_2 \lambda_2 \cdots \eta_N
\lambda_N)~. \label{ceact}
\ee
On $\ce$ there is a $\cc(\ca)$-valued Hermitian structure $\langle
\cdot,\cdot \rangle$,
\be
\langle \eta' , \eta \rangle :=
(\eta_1'^{*} \eta_1, \eta_2'^{*} \eta_2, \cdots, \eta_N'^{*} \eta_N)~ \in
~\cc(\ca)~.
\label{cherm}
\ee

To complete the geometrical construction, in addition to the algebra and
the Hilbert space we need a third element, a (generalized) Dirac
operator $D$, which, with ${\cal A}$ and ${\cal H}$ form the so called {\em
spectral triple}. The operator $D$ is self adjoint, with compact resolvent
and such that $[D,a]$ is bounded for a dense subset of the algebra, and it
is used is in the construction
of the algebra of differential forms.
These are represent as differential forms as operators on ${\cal
H}$. Define the (abstract) {\em
universal differential algebra of forms} as the $\IZ$-graded algebra
$\Omega^*{\cal A} = \bigoplus_{p\geq0}\Omega^p{\cal A}$
generated as follows:
$\Omega^0{\cal A}={\cal A}$
and $\Omega^1{\cal A}$ is generated by a set of abstract symbols $da$ linear
and which
satisfy Leibnitz rule.
Elements of $\Omega^p{\cal A}$ are linear combinations of elements of the form
\be
\omega=a_0da_1\cdots da_p
\label{pforms}
\ee

\noindent
A linear representation $\pi_D:\Omega^*{\cal A}\to{\cal B}({\cal H})$ of
the universal algebra of forms is defined by
\be
\pi_D(a_0da_1\cdots da_p )= a_0[D,a_1]\cdots [D,a_p]
\ee
Note, however, that $\pi_D(\omega)=0$ does not necessarily imply
$\pi_D(d\omega)=0$. Forms $\omega$ for which this happens are called {\em junk
forms}. They generate a $\IZ$-graded ideal in $\Omega^*{\cal A}$ and have to be
quotiented out \cite{Co1,gianni}. Then the noncommutative differential algebra
is represented by the quotient space
\be
\Omega_D^*{\cal A}=\pi_D\left[\Omega^*{\cal A}/({\rm ker}~\pi_D\oplus d~{\rm
ker}~\pi_D)\right]
\label{OmegaD}\ee
which we note depends explicitly on the particular choice of Dirac operator $D$
on the Hilbert space $\cal H$.
The algebra $\Omega_D^*{\cal A}$ determines a DeRham complex whose cohomology
groups can be computed using the conventional methods.  A discussion on
differential calculus on finite sets can be found for
example in \cite{Dimakis}.

We take $\IC^N$ for $\ch$
on which we represent elements of $\cc(\ca)$ as diagonal matrices
\be
\cc(\ca) \ni c \mapsto \mbox{diag}(\lambda_1, \lambda_2, \dots \lambda_N) \in \cb(\IC^N)
\simeq \IM_{N}(\IC)~.
\ee
Elements of $\ce$ will be realized in the same manner,
\be
\ce \ni \eta \mapsto \mbox{diag}(\eta_1, \eta_2, \dots \eta_N) \in \cb(\IC^N)
\simeq \IM_{N}(\IC)~.
\ee
Since our triple $(\cc(\ca), \ch, D)$ will be zero dimensional, the
($\IC$-valued) scalar product associated with the Hermitian structure
(\ref{cherm}) will be taken to be
\be
(\eta', \eta) = \sum_{j=1}^N \eta'^*_j \eta_j = tr \langle \eta' , \eta \rangle~,
~~~\forall ~\eta', \eta \in \ce~.
\label{csca}
\ee

By identifying $N + j$ with $j$, we take for the operator $D$, the
$N \times N$ self-adjoint matrix with elements
\be
D_{ij} = \frac{1}{\sqrt{2}\epsilon} (m^* \delta_{i+1,j} + m \delta_{i,j+1})~,
~ i,j = 1, \cdots, N~, \label{7.14.3}
\ee
where $m$ is any complex number of modulus one: $m m^* =1$.

The connection $1$-form $\rho$ on
the bundle $\ce$ is the hermitian matrix with elements
\be
\rho_{ij} =  \frac{1}{\sqrt{2}\epsilon}
(\sigma^* m^* \delta_{i+1,j} + \sigma m \delta_{i,j+1})~, \ \ \
\sigma = e^{- i \theta / N} - 1~,
~~i,j = 1, \cdots, N~.
\label{7.14.4}
\ee
One checks that, modulo junk forms, the curvature of $\rho$ vanishes,
\be
d \rho + \rho^2 = 0~. \label{7.14.5}
\ee
It is also possible to prove that $\rho$ is a `pure gauge' for
$\theta = 2 \pi k$, with $k$~ any integer, that is
that there exists a $c \in \cc(\ca)$ such that $\rho = c^{-1} d c$.
If $c = \mbox{diag}(\lambda_1 ,\lambda_2,
\ldots ,\lambda_N )$, then any such $c$ will be given by $\lambda_1 =
\lambda~, ~\lambda_2 =
e^{i 2\pi k / N } \lambda~, ... , ~\lambda_j = e^{i 2\pi k (j-1) / N } \lambda~, ...,
~\lambda_N = e^{i 2\pi k (N - 1) / N } \lambda$, with $\lambda$ not equal to $0$
(these properties are the analogues of the properties of the
connection $i \theta d x$ in the `continuum' limit).

The covariant derivative $\nabla_\theta$ on $\ce$, $\nabla_\theta : \ce \ra \ce
\otimes_{\cc(\ca)} \Omega^1(\cc(\ca))$  is then given by
\be
\nabla_\theta \eta = [D, \eta] + \rho \eta~, ~~~ \forall ~\eta \in \ce~.
\label{ccovder}
\ee
In order to define the Laplacian $\Delta_\theta$ one first introduces a `dual'
operator
$\nabla_\theta^*$ via
\be
(\nabla_\theta \eta', \nabla_\theta \eta) = (\eta', \nabla_\theta^{*}
\nabla_\theta
\eta)~,  ~~~\forall ~\eta', \eta \in \ce. \label{cdcovder}
\ee
The Laplacian $\Delta_\theta$ on $\ce$, $\Delta_\theta : \ce \ra \ce$, can then
be defined by
\be
\Delta_\theta \eta = - q (\nabla_\theta)^{*} \nabla_\theta \eta ~,~~~ \forall
~\eta
\in
\ce~,
\label{clap}
\ee
where $q$ is the orthogonal projector on $\ce$ for the scalar product
$( \cdot, \cdot )$ in (\ref{csca}). This projection operator is readily seen
to be given by
\be
(q M)_{ij} = M_{ii}\delta_{ij}~,~~~\mbox{no summation on}~i~,
\label{7.14.7}
\ee
with $M$ any element in $\IM_{N}(\IC)$. Hence, the action of $\Delta_\theta$ on
the element $\eta = (\eta_1, \cdots, \eta_N)$~, $\eta_{N+1} = \eta_1$, is
explicitly given by
\bea
 (\Delta_\theta \eta)_{ij} &=& - (\nabla_\theta^{*} \nabla_\theta \eta)_{ii}
\delta_{ij}~, \nonumber \\
- (\nabla_\theta^{*} \nabla_\theta\eta)_{ii} &=&
\left\{ - \left[ D, [D, \eta ] \right] - 2 \rho [D,
\eta] - \rho^2 \eta \right\}_{ii} \nonumber \\
&=& \frac{1}{\epsilon^2}
\left[ e^{-i\theta/N} \eta_{i-1} - 2 \eta_i +  e^{i\theta/N} \eta_{i+1} \right]~;
~~~i = 1, 2, \cdots , N~.
\label{7.15}
\eea
The associated eigenvalue problem
\be
\Delta_\theta \eta = \lambda \eta ~,
\ee
has solutions
\bea
\lambda &=& \lambda_k = \frac{2}{\epsilon ^2}
\left[\cos (k + \frac{\theta}{N}) -1 \right]~, \label{ceigen} \\
\eta &=& \eta^{(k)}
= \mbox{diag}(\eta^{(k)}_1, \eta^{(k)}_2, \cdots, \eta^{(k)}_N)~,
~~~k = m \frac{2\pi}{N}~,~ m = 1, 2, \cdots, N~,
\eea
with each component $\eta^{(k)}_j$ having an expression of the form
\be
\eta^{(k)}_j = A^{(k)} e^{i kj} + B^{(k)}e^{-i kj}~,~~~
A^{(k)}, B^{(k)}\in \IC~.
\ee
The eigenvalues (\ref{ceigen}) are an approximation to the
continuum answers $-4k^2~, ~k \in \IR$.

\section{Conclusions}

In this note we described a way to look at manifolds based on a coarse
approximation which however retains the principal topological
characteristics of the original space. The motivations for this work have
been in the approximation processes natural in physics: finite size of the
detectors, impossibility to probe very short distances, etc.

We would like to note also that, apart from the
measurement problems, there is a scale (Planck Scale) at which the
structure of space time is very likely not to be describable by the usual
tools of geometry, and in this case Noncommutative Geometry seems to be the
ideal tool for a more general description of spacetime. Some attempts on
considering spacetime at the Planck scale as composed of noncommutative
objects, finite or with a limited number of degrees of freedom, have been
done for example in \cite{Heller,Requardt,Zapatrin}. Another fruitful arena
in which similar concepts play an important role, and which recently is
having interesting "contaminations" with noncommutative geometry,
is certainly string theory, the discussion of which will take too much
room, and which we omit here.

In conclusion, to abandon the usual concepts (prejudices?) of a geometry
made of separable points, lines, and complex functions is not only an
important step for pure mathematics, but in a few years we may come to
consider it as the most natural step imposed by physics.

\bigskip
\bigskip

\noindent
{\bf Acknowledgements.} \\
The material presented here is based on work done in
collaborations with A.P.~Balachandran, G.~Bimonte, E.~Ercolessi, G.~Sparano,
P.~Teotonio-Sobrinho. We thank C.~Buzzanca, D.~Kastler and M.~Rosso
for their kind invitation to partecipate to this nice adventure.
%
%
\bibliographystyle{unsrt}

\end{document}